\documentclass[floatfix,twocolumn,amsmath,amssymb,superscriptaddress]{revtex4-2}

\usepackage{graphicx}
\usepackage{hyperref}
\usepackage{braket}
\usepackage{bbm}
\usepackage[utf8]{inputenc}

\usepackage[svgnames]{xcolor}
\usepackage[normalem]{ulem}
\input{macros.sty}

\begin{document}

\title{Optimal Cooling of Multiple Levitated Particles: Theory of Far-Field Wavefront-Shaping}

\author{Jakob Hüpfl}
\affiliation{Institute for Theoretical Physics, Vienna University of Technology (TU Wien), Vienna, Austria}
\author{Nicolas Bachelard}
\affiliation{Institute for Theoretical Physics, Vienna University of Technology (TU Wien), Vienna, Austria}
\affiliation{CNRS, LOMA, UMR 5798, Université de Bordeaux, Talence, France}
\author{Markus Kaczvinszki}
\affiliation{Institute for Theoretical Physics, Vienna University of Technology (TU Wien), Vienna, Austria}
\author{Michael Horodynski}
\affiliation{Institute for Theoretical Physics, Vienna University of Technology (TU Wien), Vienna, Austria}
\author{Matthias Kühmayer}
\affiliation{Institute for Theoretical Physics, Vienna University of Technology (TU Wien), Vienna, Austria}
\author{Stefan Rotter}
\email{stefan.rotter@tuwien.ac.at}
\affiliation{Institute for Theoretical Physics, Vienna University of Technology (TU Wien), Vienna, Austria}

\begin{abstract}
The opportunity to manipulate small-scale objects pushes us to the limits of our understanding of physics. 
Particularly promising in this regard is the interdisciplinary field of levitation, in which light fields can be harnessed to isolate nano-particles from their environment by levitating them optically.
When cooled down towards their motional quantum ground state, levitated systems offer the tantalizing prospect of displaying mesoscopic quantum properties.
Currently restricted to single objects with simple shapes, the interest in levitation is currently moving towards the manipulation of more complex structures, such as those featuring multiple particles or different degrees of freedom.
Unfortunately, current cooling techniques are mostly designed for single objects and thus cannot easily be multiplexed to address such coupled many-body systems. 
Here, we present an approach based on the spatial modulation of light in the far-field to cool down multiple nano-objects in parallel. 
Our procedure is based on the experimentally measurable scattering matrix and on its changes with time. 
We demonstrate how to compose from these ingredients a linear energy-shift operator, whose eigenstates are identified as the incoming wavefronts that implement the most efficient cooling of complex moving ensembles of levitated particles. 
Submitted in parallel with \cite{Arxiv:2103.12592}, this article provides a theoretical and numerical study of the expected cooling performance as well as of the robustness of the method against environmental parameters. 
\end{abstract}
\maketitle
\section{Introduction}
With the ability to simultaneously manipulate multiple objects comes the possibility of interrogating the basics of fundamental physics. 
For instance, Maxwell's suggested rearrangement of a gas of particles by a demon \cite{PhysRevX.7.021051} has questioned our understanding of the second law of thermodynamics.
Many-body manipulation also enables the assembly of new states of matter. 
For instance, in quantum physics, the cooling of collective systems towards their ground state has been acknowledged by multiple Nobel prizes, e.g., for laser cooling \cite{PhysRevA.20.1521} or the realization of Bose-Einstein condensates \cite{PhysRevLett.80.2027}. 
Yet, conventional approaches in many-body manipulation are typically restricted to small scales and often prove to be case-specific. 
At the mesoscopic scale, the field of levitation has recently emerged as an opportunity to harness optical forces to manipulate objects decoupled from their environment \cite{1,science.abg3027}, while offering remarkable opportunities for high-resolution sensing \cite{2,3,4} or non-equilibrium thermodynamics \cite{5,6}. 
Mostly implemented through optical tweezers, levitation has been mainly restricted to single objects so far; since the multiplexing of traps reveals challenging due to the optical binding amongst elements \cite{19,RevModPhys.82.1767}.

Experimentally, levitated objects are manipulated through schemes relying on local information. 
For instance, in cavity cooling, an element is positioned (e.g., with tweezers) at the peak of the local density of optical states inside a cavity where it will be cooled down by radiation pressure \cite{9,10,11}. 
Through a modified cavity-cooling scheme known as coherent scattering \cite{12,13,14}, the ground state of a nanometer-size bead was recently reached \cite{10.1126/science.aba3993}. 
In feedback cooling \cite{PhysRevLett.109.103603}, the position of a trapped particle is constantly monitored and processed by an electronic control loop, which modulates the involved tweezers to produce a force that continually opposes the object’s motion and cools it down to its ground state \cite{10.1038/s41586-021-03617-w,magrini2021real}. 
The efficiency of such schemes is ultimately limited by the ability to impose or extract local features: cavity cooling requires high-quality resonances and a precise control over the particle’s location \cite{10.1126/science.aba3993}, while feedback cooling relies on the detection of motion and suffers from calibration issues \cite{18}. 

Yet, the same electromagnetic field that produces optical forces at the particle also transfers information on this object to the far-field when being scattered away from its target. 
In complex and multi-element systems, the information on how incoming and outgoing scattering states are related, is conveniently stored in the scattering matrix, which is routinely measured through wavefront shaping \cite{PhysRevLett.104.100601,RevModPhys.89.015005}. 
Such matrices provide access to tailor-made light states for applications ranging from bio-imaging \cite{22} to quantum optics \cite{defienne2016two} and have recently been suggested for probing structural features like mechanical actions \cite{24,horodynski2020optimal} or optimal sensing \cite{26}. 
The question we will ask here is whether such far-field scattering information can also be used to develop new control schemes for the cooling of many-body systems.

The affirmative answer we provide here will allow us to present for the first time (to the best of our knowledge) an optimal procedure to cool down levitated many-body systems through wavefront shaping. 
The cooling of several objects in parallel is achieved through a spatial modulation (in the far field) of an incoming light field, which is designed to exert (on average) optical forces counteracting the instantaneous motion of each particle. 
Our paper is organized as follows: in section \ref{sec:coolingprocedure}, we derive the cornerstone relation of this work (the so-called energy-shift relation) and explain the procedure to extract the wavefronts performing the cooling. 
Then, in section \ref{sec:coolingsimulations}, the procedure is applied to cool down many-body systems composed of spherical and non-spherical nano-particles alike, which are either freely moving or optically trapped. 
At last, in section \ref{sec:discussion}, we discuss the robustness and the performance of the method. 
This article is submitted in parallel to \cite{Arxiv:2103.12592}, which provides a considerably shortened account of the main results presented here. 

\section{Cooling procedure}
\label{sec:coolingprocedure}
\begin{figure*}
	\begin{tabular}{l{0.01\textwidth} f{0.48\textwidth} l{0.01\textwidth} f{0.48\textwidth}}
	(a)& %
	\includegraphics[scale=0.4, valign=t]{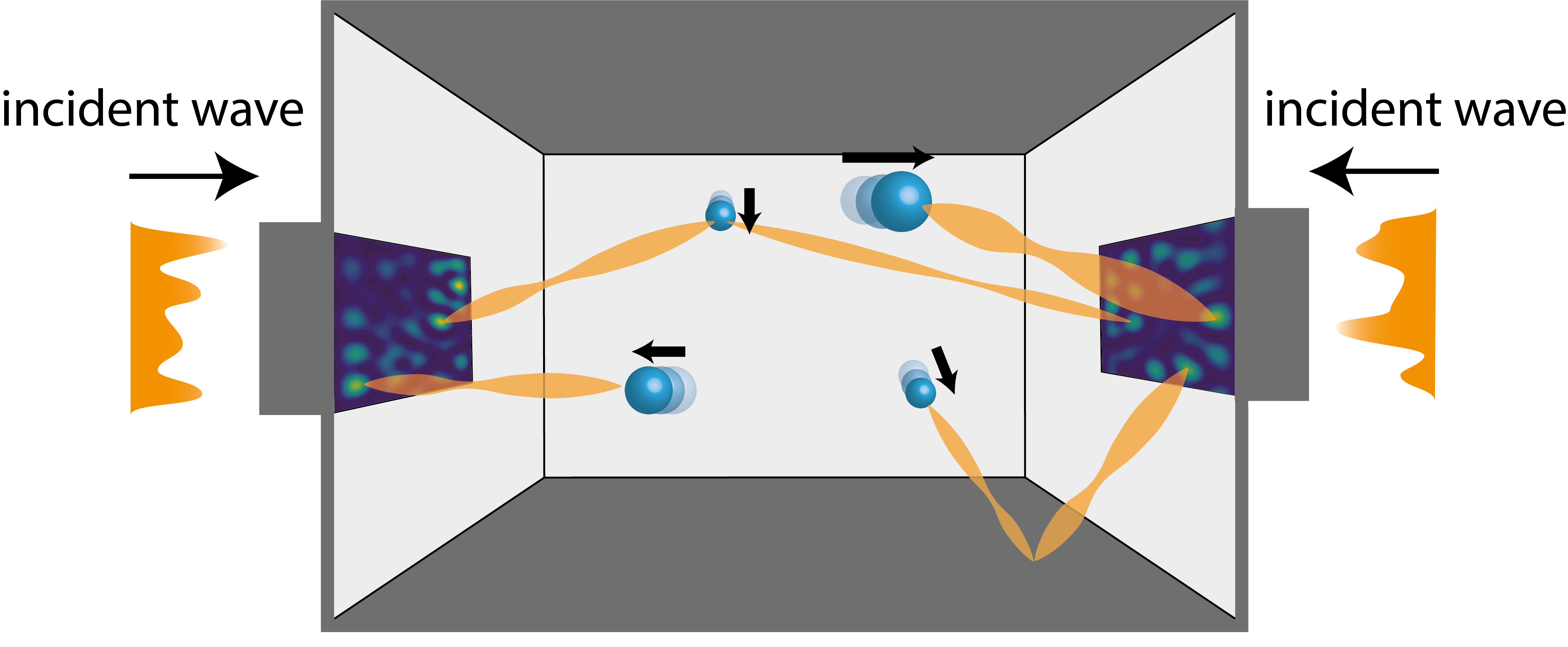}&%
    (b)&%
    \hspace*{-20pt}
	\includegraphics[scale=0.66, valign=t]{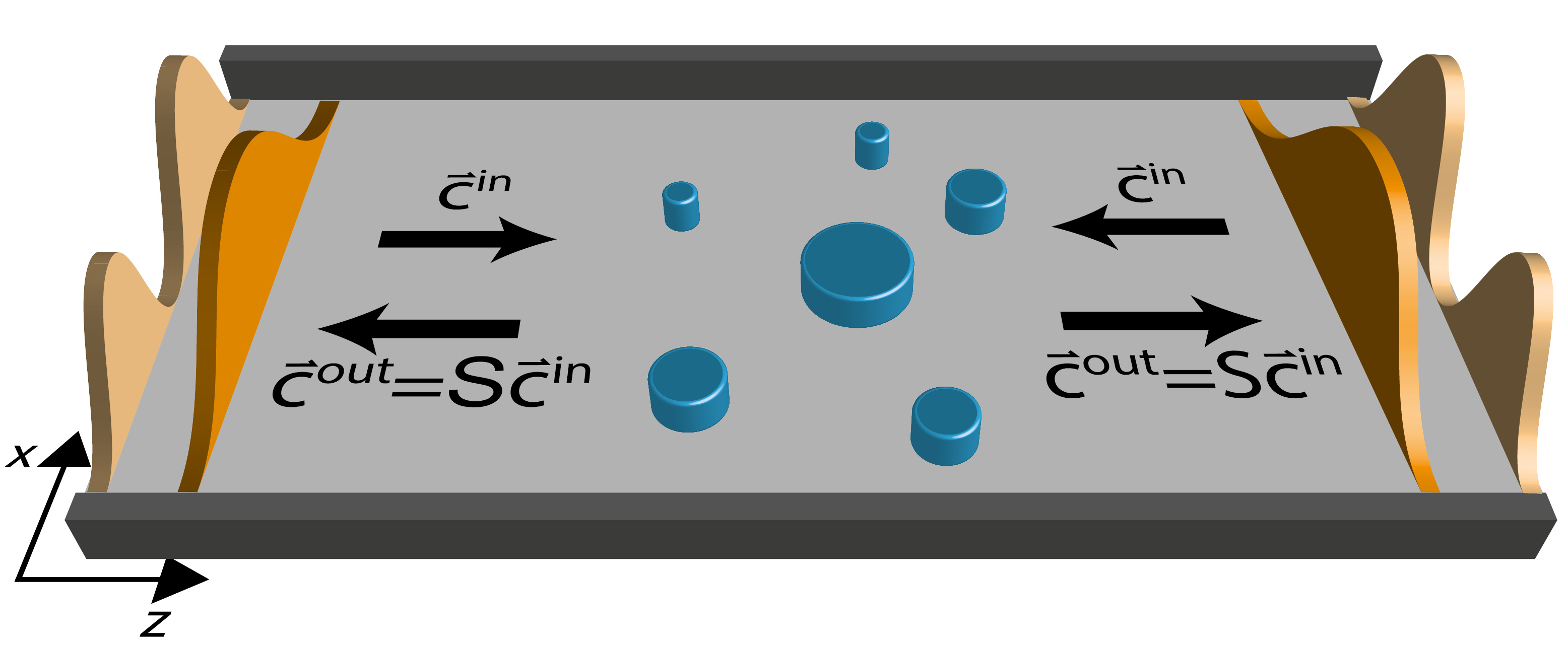}
	\end{tabular}
	\caption{(a) Simplistic illustration of the cooling concept. A many-body system is composed of nano-particles (blue beads) in random motion (black arrows). 
	The system is opened on two sides, through which spatially-modulated light fields (``incident wave'', blue to yellow patterns) get injected. 
	The wavefronts (orange wave lines) are designed to counteract the motion of the nano-particles. 
	(b) In our numerical simulations, this cooling approach is implemented in a two-dimensional system, such as a thin waveguide (grey structure, small extension in $y$-direction), in which nano-particles (blue cylinders) experience a 2D motion in the ($x$,$z$) plane. 
	The shaped wavefronts, which are used to cool the motion of the nano-particles, get injected from both sides. 
	Inside the waveguide, the field is composed of propagating transverse electric modes (the intensity distribution of the first two modes are indicated by orange shapes).
	The shaped incident and outgoing wavefronts are expressed as a linear combination of such modes through the coefficient vectors $\vec{c}^{\,\text{in}}$, $\vec{c}^{\,\text{out}}$. 
	These input and output vectors are connected through the scattering matrix, $\matrix{S}(t)$, which evolves through time $t$ due to particles' motion.}
	\label{fig:3D}
\end{figure*}
We consider the general setting of an electromagnetic field propagating in a dielectric medium, which is described by a linear permittivity, $\epsilon$, and a linear permeability, $\mu$. 
The field is made up of time-harmonic waves oscillating at an optical frequency $\omega$, while the medium consists of a many-body system composed of $N_{\mathrm{scat}}$ moving nano-particles with $\Ndof$ total degrees of freedom. 
Here, we aim to cool this complex system through the injection of spatially modulated light fields [Fig. \ref{fig:3D}(a)]. 
The motion of the nano-particles as a function of time $t$ happens on a time-scale that is much longer than that of the light field (i.e., the particle velocity is small compared to the speed of light, $v \ll c$). 
This adiabatic motion is included in our theory by a non-uniform and time-dependent permittivity, $\epsilon(r,t)$, that is slowly evolving over time with the displacements of the rigid particles. 
The far-field scattering of these nano-particles is encapsulated in a linear scattering matrix, $\S$. 
Physically, the scattering matrix relates any field (i.e., wavefront) incoming into the system, $\vec{c}^{\,\text{in}}$, to the field that is scattered towards the far field, $\vec{c}^{\,\text{out}}=\S\vec{c}^{\,\text{in}}$ \cite{RevModPhys.89.015005}. 
Experimentally, $\S$ can be inferred by sending in a succession of predefined spatially-modulated wavefronts, while recording the corresponding scattered fields \cite{PhysRevLett.104.100601}. 
To access the total (i.e., mechanical) energy of the nano-particles, we recast $\S$ into a new operator, referred to as the energy-shift matrix, 
\begin{equation}
    \Qt = - \i \S^\dagger \der[t] \S\,,
\end{equation}
where $\der[t]$ corresponds to the time derivative due to the motion of the particles.
In the context of electron transport, a modified version of this operator was introduced by Avron \textit{et al.}~to describe how externally driven charge pumps pass electrons through a conductor \cite{avron2002time,moskalets2004adiabatic}. 
Here, instead, for a system subject to weak non-conservative forces, we will explain how $\Qt$ can be harnessed to construct spatially-modulated wavefronts that are able to optimally reduce the macroscopic total energy of many-body systems--which is equivalent to a cooling of their center-of-mass temperature \cite{10.1088/13616633/ab6100}. 
Specifically, we will introduce a new cooling strategy, in which wavefronts are designed in real time to exert optical forces that are ``optimally'' counteracting the motion of all the nano-particles simultaneously.

\subsection{Energy-shift relation}
\label{sec:GWSrelation}
We will now give a sketch of the main analytic results of this article. 
The complex amplitudes of the time harmonic optical fields are governed by Maxwell's equations,
\begin{equation}
\label{eq:PDE:Efield}
\begin{split}
\rot \E &= \i\omega \vec{B},\\
\rot \H &= -\i\omega \vec{D},
\end{split}
\end{equation}
where the different components implicitly depend on the nano-particles' positions and therefore evolve with time $t$. 
For the sake of simplicity, we will focus in the main text on dielectric elements (i.e., described by a real time-dependent permittivity $\epsilon(r,t)$ and constant permeability $\mu=\mu_0$). 
We show in appendix \ref{app:GWSrelation} that, in a scattering region $\Omega$ of boundary $\partial \Omega$ with an outgoing normal vector $\vec{n}$, the following equality holds: 
\begin{equation}
\begin{split}
\int_{\partial \Omega} (\E^* \times \der[t]\H - \H^* \times \der[t]\E) \cdot \d \n = \i \omega\int_{\Omega} |\E|^2  \der[t] \epsilon\,.
\end{split}
\label{eq:EnergyConservation}
\end{equation}
We emphasize here that the time derivative $\der[t]$ is in relation to the evolution of the particles and not to the light field itself.
What makes the above Eq.~(\ref{eq:EnergyConservation}) particularly useful is the fact that its left-hand side only involves the fields at the boundary of the scattering region, while its right-hand side describes the change in the particles' total energy $E_{\text{tot}}$ imposed by light forces. 
Recasting this relation in a form involving the energy-shift operator results in
\begin{equation}
\label{eq:GWSrelation}
	\c^{\,\text{in},\dagger} \Qt \c^{\,\text{in}} = 2 \omega (\der[t] E_{\text{tot}} - P_{\text{nc}})\,,
\end{equation}
where $\c^{\,\text{in}}$ describes an incoming spatially-shaped wavefront and $P_{\text{nc}}$ the power of other non-conservative forces (e.g., viscous friction). 
Here, we stress that $E_{\text{tot}}$ comprises the macroscopic mechanical energy of all the nano-particles and thus encompasses the kinetic energy of both translational and rotational degrees of freedom as well as any potential energy (e.g., trapping potential). 

The energy-shift relation of Eq.~(\ref{eq:GWSrelation}) constitutes the cornerstone of this paper as it serves as the starting point for the implementation of our approach to many-body cooling. 
When non-conservative contributions (i.e., $P_{\text{nc}}$) remain weak, Eq.~(\ref{eq:GWSrelation}) reduces to $\c^{\,\text{in},\dagger} \Qt \c^{\,\text{in}} \approx 2 \omega \der[t] E_{\text{tot}}$. For loss-less scattering systems $\S$ is unitary, resulting in $\Qt$ being Hermitian with a corresponding decomposition into an orthogonal eigenbasis with real eigenvalues. %therefore featuring an easily accessible real-eigenvalue decomposition. 
Thus, at any time $t$, the eigenvector of $\Qt$ associated with the smallest (i.e., most negative) eigenvalue will perform an instantaneous reduction of $E_{\text{tot}}$ that is optimal. 
In the rest of this article, we will refer to such eigenvectors as ``optimal cooling states''. 
While in real-world experiments the scattering matrix is typically not measured entirely and non-conservative effects contribute, we will see in section \ref{sec:missing}, that even in such realistic scenarios a cooling scheme based on Eq.~\ref{eq:GWSrelation} still manages to display strong cooling performance.

\subsection{Cooling procedure}
\label{seq:coolingprocedure}
For the implementation of our cooling procedure, we suggest a stroboscopic measurement of the scattering matrix $\S(t)$ with a sampling time step $\coolstep$. 
At a certain time $t-\coolstep$, a basis of non-perturbing test states with weak power is first applied successively to the nano-particles and the corresponding outgoing fields are measured to extract $\S(t-\coolstep)$. 
Then, an approximation of the energy-shift operator at time $t$ is obtained with two consecutive measurements of $\S$, taking the following form: $\Qt(t) \approx - \i \S^{\dagger}(t) [\S(t)-\S(t-\coolstep)]/\coolstep$. 
Finally, in line with Eq.~(\ref{eq:GWSrelation}), the optimal cooling state is found by computing the eigenstate of $\Qt(t)$ associated with its most negative eigenvalue. 
This state is then applied at a strong power to the system during $\Delta t_{\cool}$ to cool the system down before the whole procedure is repeated at time $t+\Delta t_{\cool}$. 

\subsection{Stochastic motion of the nano-particles}
\label{sec:model}
Throughout this manuscript, our method will be applied to nano-objects with various geometries that are in contact with a thermal bath (temperature $\tempenv$) and subject to Brownian motion. 
For instance, under partial-vacuum conditions, the dynamic of a nano-sphere of mass $\mass_i$ (with only translational degrees of freedom) is governed by a Langevin equation,
\begin{equation}
\label{eq:langevin_translation}
\dot{\momvec}_i(t) = - \gamma_i \momvec_i(t) + m_i \vec{g} + \forcevec_{\cool,i}(t) + \sqrt{2 D_i}\, \whitenoise_i(t),
\end{equation}
in which $\momvec_i$ stands for the particle's momentum, $\gamma_i$ for its damping rate and $\vec{g}$ for gravity, while $\forcevec_{\cool,i}$ describes the force of the light field. 
The Brownian motion induced by the coupling to the thermal bath manifests itself in Eq.~(\ref{eq:langevin_translation}) through the presence of a stochastic term $\sqrt{2 D_i}\, \whitenoise_i(t)$, whose amplitude is set by the dissipation-fluctuation relation $D_i = \kB \tempenv \gamma_i \mass_i$ and in which $\whitenoise_i(t)$ corresponds to an independent Gaussian white noise with unitary variance. 
A complete description of the dynamics of non-spherical objects (with both translational and rotational degrees of freedom) is provided in appendix \ref{app:complete_description}.
 
%% Figure 2
\begin{figure*}[t!]%76.8 x 12.8
	\begin{tabular}{l{0.01\textwidth} f{0.31\textwidth} l{0.01\textwidth} f{0.31\textwidth} l{0.01\textwidth} f{0.31\textwidth}}
	(a)& %
	\centering 
	\includegraphics[scale=0.65, valign=t]{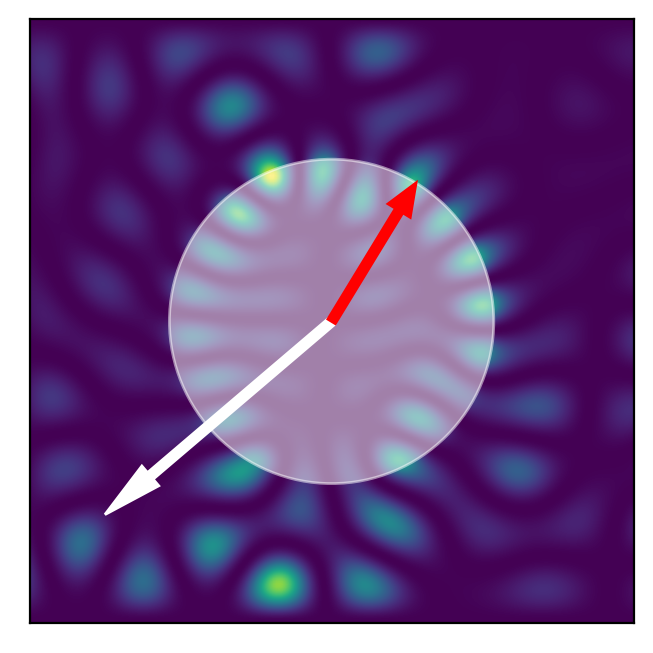}
    $\corr_w= \corr_w^{\text{trans}} = - 0.95$&%
    (b)&%
	\centering 
	\includegraphics[scale=0.65, valign=t]{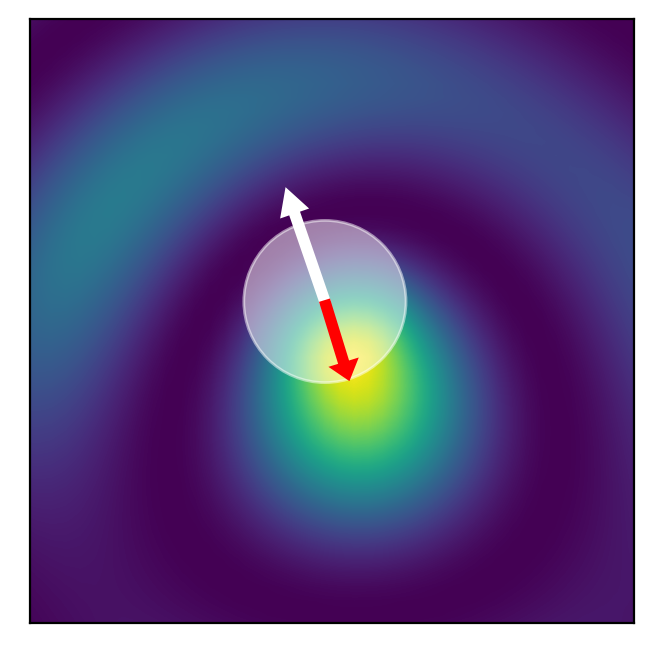}
    $\corr_w= \corr_w^{\text{trans}} = - 0.999$&
    (c)&%
	\includegraphics[scale=0.65, valign=t]{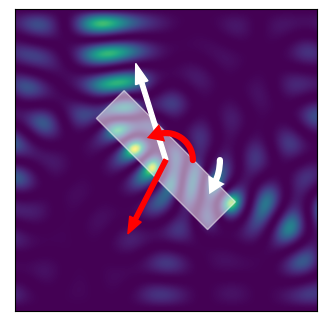}
    $\corr_w = - 0.6$
	\end{tabular}
	\caption{(a),(b) Optimal cooling states applied to a single circular nano-bead with radius $750 $~nm in (a) and $75 $~nm in (b), respectively. 
	The waveguide in which the particles move has a width of $W=5.5 $~$\mu$m, corresponding to $\modes=20$ propagating optical modes. 
	The instantaneous velocities of the beads are marked by white arrows, while the optimal cooling states create scattering patterns that produce forces marked by red arrows. 
	The quantity $\corr_w^{\text{trans}}$ below the panels characterizes the correlation between the velocity and the force vectors (limited by the globally optimum value of $\corr_w^{\text{trans}} =-1$). 
	(c) Optimal cooling states applied to a rectangular particle experiencing translation and rotation. 
	The waveguide dimensions are identical to (a) and (b), while the longer edge of the rectangle is set to $\approx 1.5 $~$\mu$m. 
	The rigid body translation and rotation of the particle are marked by the straight and curved white arrows, respectively. 
	The optimal cooling state induces a force on the center-of-mass and a torque that are marked by the straight and curved red arrows. 
	$\corr_w$ is the total correlation of center-of-mass and angular velocity against the optical force and torque.}
	\label{fig:focus}
\end{figure*}
 
\section{Simulations}
\label{sec:coolingsimulations}
\subsection{Numerical model}
\label{sec:parameters}
To keep our numerical simulations computationally manageable, we will restrict ourselves to nano-particles following a 2D motion in the $(x,z)$ plane. 
Moreover, to reduce the number of optical modes and thus to limit the dimension of $\S$, we confine the nano-particles into a waveguide with a thin rectangular cross section, where also the electromagnetic field can be treated as scalar. 
A sketch of this can be seen in Fig.~\ref{fig:3D}(b). 
Yet, we emphasize that this restriction to a 2D geometry is only intended to reduce computational complexity. 
As we show in appendix \ref{app:GWSrelation}, our method can also be applied in free space as well as for 3D objects with non-trivial shapes. 
The field propagation can be directly linked to the scattering problem of a two dimensional waveguide filled with circular scatterers \cite{horodynski2020optimal}. 
Modulated wavefronts get injected from both sides, while the motion of the nano-particles fulfills Eq.~(\ref{eq:langevin_translation}) and the rare occurrences of collisions between particles or with the waveguide walls are handled through an elastic model preserving kinetic energy. 
Moreover, the average displacement of the objects remains small compared to the system's dimensions, which makes the bouncing of the particles off the walls rare events.  

Inside the thin waveguide only TE waves can propagate and the electric field reads $\E^{\,\text{TE}} = \psi(x,z) \hat{e}_y$, while the scatterers are $y$-symmetric dielectrics ($\nablavec \epsilon \perp \E^{\,\text{TE}}$). 
Thus, at each time $t$, the field propagation fulfills the scalar Helmholtz equation, 
\begin{equation}
(\Delta + \omega^2 \mu \epsilon(x,z)) \psi(x,z)  = 0\,.
\label{eq:Helmholtz}
\end{equation}
The electric field inside the waveguide will be decomposed into a discrete set of transverse propagating modes with spatial profile $\vec{\eta}_j(x)$ and corresponding wave vector in propagation direction $k_{z,j}$ (orange shapes in Fig.~\ref{fig:3D}(b)). 
These flux-normalized eigenmodes of Eq.~(\ref{eq:Helmholtz}) in the asymptotic (scattering-free) part of the waveguide, serve as a complete basis to assemble $\S$. 
Equivalently, the incoming modulated wavefronts in Eq.~(\ref{eq:GWSrelation}) will be described as a linear combination of such modes, $\sum_j \vec{\eta}_j(x,y)c_{j}^{\text{in}}e^{\i k_{z,j} z}$, in which the coefficients $c_{j}^{\text{in}}$ constitute the vector $\c^{\,\text{in}}$.

If not stated otherwise, our simulations will be run for a waveguide width $W \approx 2.8 $~$\mu$m, corresponding to $\modes=10$ propagating modes for an optical wavelength $\lambda = 532 $~nm. 
The particles are coupled to a thermal bath at a temperature $\tempenv=30$~K, providing a damping rate $\gamma = 6$~Hz  (see section \ref{sec:model}). 
Before cooling, the nano-objects will start from individual velocities following a Boltzmann distribution with a mean kinetic energy $\kB \tempenv/2$ along each degree of freedom. 
The sampling time will be fixed to $\coolstep = 1$~$\mu$s. 
More details about the simulations' protocol are provided in appendix \ref{app:implementation}.

%% Figure 3
\subsection{Single freely-moving objects}
\label{sec:rotation}
\begin{figure*}
	\begin{tabular}{l{0.01\textwidth} f{0.48\textwidth} l{0.01\textwidth} f{0.48\textwidth}}
	(a)&%
		\hspace*{-3pt}\includegraphics[scale=1,valign=t]{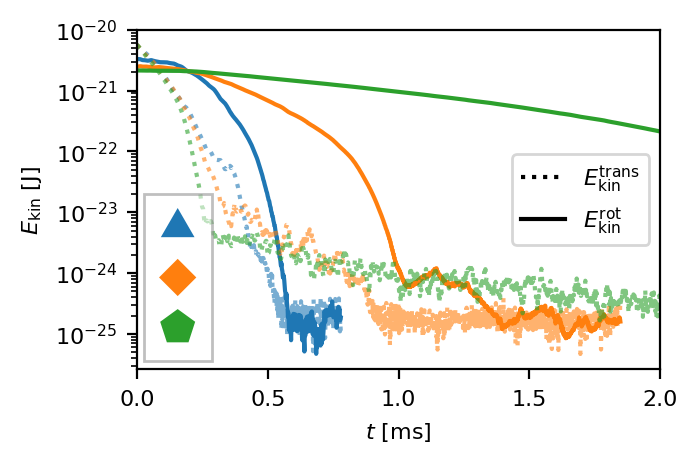}&%
    (b)&%
        \vspace*{2pt}
        \hspace*{17pt}
		\includegraphics[scale=0.16,valign=t]{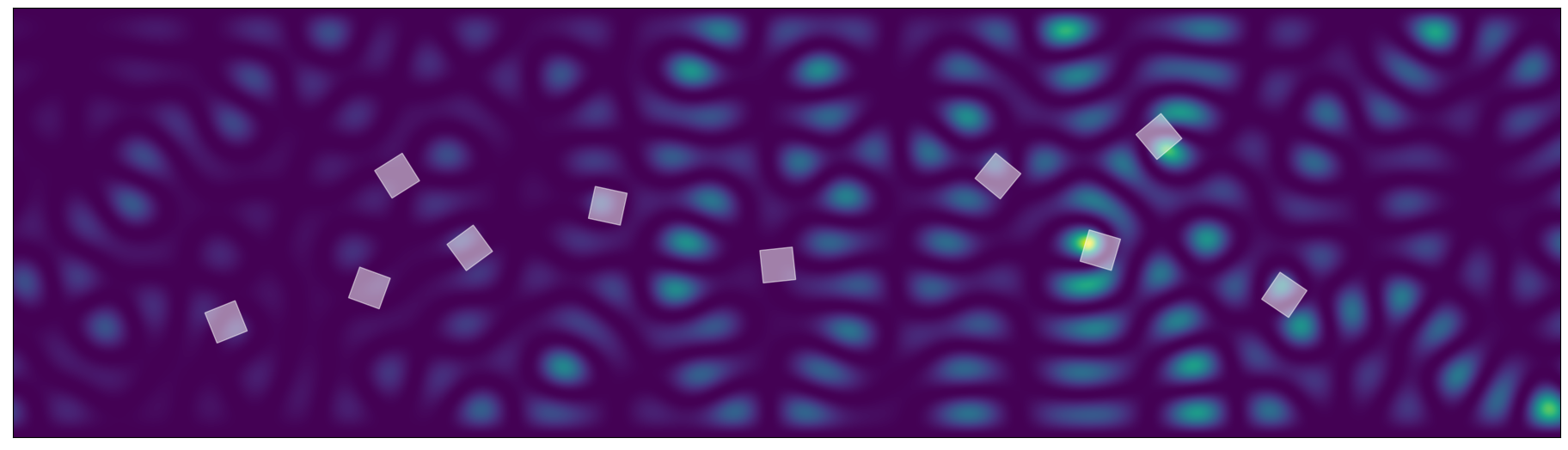}
		\hspace*{-5pt}\raisebox{8pt}[0pt][0pt]{\makebox[\columnwidth][c]{\includegraphics[scale=1.1,valign=t]{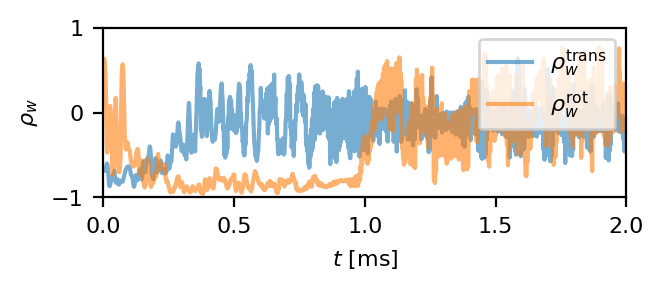}}}%
		\tabularnewline
	(c)&%
	\multicolumn{3}{f{0.96\textwidth}}{
    \hspace*{-0.01\textwidth}
		\includegraphics[scale=1,valign=t]{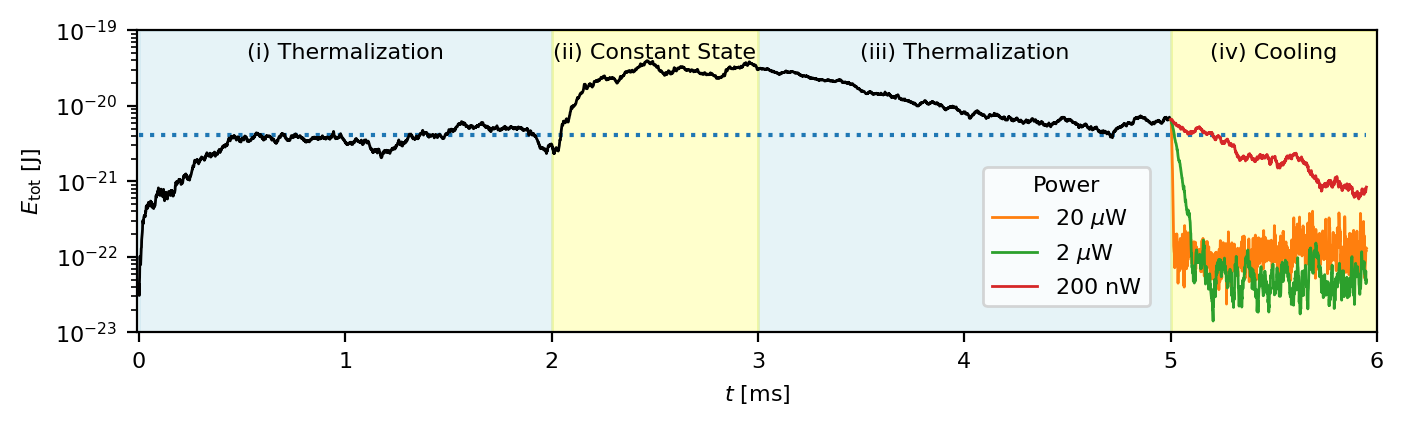}}
	\end{tabular}
	\caption{
	(a) Results of the cooling procedure applied to 10 triangles, 10 squares and 10 pentagons, respectively, moving in a waveguide without friction (all particles have the same outer radius of $r = 150 $~nm). 
	The colors blue, orange and green indicate the particle geometry (see inset). 
	Dotted and solid curves display the evolution of the translational, $E_{\kin}^{\text{trans}}$, and rotational, $E_{\kin}^{\text{rot}}$, kinetic energy throughout the process. 
	The sampling time $\coolstep = 1$~$\mu$s and the light field is characterized by a power $P= 20 $~$\mu$W, a wavelength $\lambda = 532 $~nm and $\modes = 10$ transverse waveguide modes. 
	Compared to translational degrees of freedom, rotational degrees of freedom are addressed at characteristic times dependent of the particles' shapes. 
	(b) Evolution of the correlation between force-velocity ($\corr_w^{\text{trans}}$, blue) and between torque-angular velocity ($\corr_w^{\rot}$, orange) during the cooling of the 10 squares, see orange curves in (a) (data includes a running average over 10 time steps). 
	At early times, only the translation degrees of freedom are cooled down and $\corr_w^{\text{trans}}$ settles to high negative values while $\corr_w^{\rot}$ remains low. 
	The situation is reversed when the rotational degrees of freedom start to be cooled down. 
	A snapshot of the optical field (yellow to blue pattern) produced by an optimal cooling state is provided in inset. 
	(c) Simulation results for $\Nscat=10$ beads of radius $75 $~nm, subject now to a stochastic motion as in  Eq.~(\ref{eq:langevin_translation}) that is characterized by a damping rate $\damping = 1$~kHz and a surrounding temperature $\tempenv = 30$~K. 
	All the other parameters are identical to the ones used in (a) and (b). 
	In phase (i), the velocities of the beads start out thermally distributed according to a (low) initial temperature of $T=0.01 \cdot \tempenv$. 
	Thermalization to the bath temperature at $\tempenv = 30$~K is observed in the first $\approx 2$~ms (blue dotted line indicates the prediction from the equipartition theorem). 
	Over the next $\approx 1$~ms in the constant state phase (ii), a field constant over time ($P= 20$~$\mu$W) gets injected from both waveguide leads and produces an increase in energy. 
	Afterwards, the field is turned off and in phase (iii), from $3$ ms to $5$ ms, the particles thermalize back towards $\tempenv$. 
	Finally, in the cooling phase (iv), our method is applied using a power of $20$~$\mu$W, $2$~$\mu$W or $200$~nW (orange, green and red curves, respectively), leading to a significant drop in energy.}
	\label{fig:rot}
\end{figure*}
To illustrate the role of optimal cooling states, we first consider simple single-particle systems. 
In Fig.~\ref{fig:focus}, our method is applied to different individual objects. 
At a given time $t$, the optimal cooling state is computed following the strategy described in section \ref{seq:coolingprocedure} and injected into the waveguide. 
We observe that optimal cooling states generally correspond to complex light fields, which produce optical forces (red arrows) that are systematically opposed to the instantaneous velocities of the nano-particles (white arrows).
For translational degrees of freedom (labeled as ``trans'' in the following), the instantaneous cooling efficiency can be quantified through a weighted force-velocity correlation. 
For multiple particles labelled by the index $i$, this correlation is defined as,
\begin{equation}
\label{eq:focus}
\corr_w^{\text{trans}} \coloneqq \frac{\sum_i \momvec_i \cdot \forcevec_{\cool,i} (\mass_i)^{-1}}{\sqrt{\sum_i \momvec_i^{\,2} (\mass_i)^{-1}}\sqrt{\sum_i \forcevec_{\cool,i}^2 (\mass_i)^{-1}}}\,,
\end{equation}
which reduces to $\corr_w^{\text{trans}}=\momvec\cdot\forcevec/(\|\momvec\| \|\forcevec\|)$ for single particles.
Physically, an ideal cooling will display a maximal anti-correlation (i.e., $\corr_w^{\text{trans}} = -1$). 
In the case of multiple and identical particles, $\corr_w^{\text{trans}}$ measures the co-linearity between the force and velocity vectors. 
For spherical objects that are only provided with translational degrees of freedom, Figs.~\ref{fig:focus}(a) and (b) show that the optimal cooling state applies a force that is almost completely anti-correlated to the motion of the particles, both with $r \gg \lambda$ and $r \ll \lambda$, respectively.
In appendix \ref{app:torquerotation}, we explain that Eq.~(\ref{eq:focus}) can be extended to include both translational and rotational degrees of freedom (the latter will be labeled as ``rot'' in the following).
Figure \ref{fig:focus}(c) provides an example where the optimal cooling state is able to counteract both rotational and translational motions at the same time. 
Here, a correlation of $\corr_w=-0.6$ is reached and will enable to efficiently slow down all the degrees of freedom (example of efficient cooling provided in Fig.~\ref{fig:rot}(b)).   

\subsection{Multiple freely-moving objects}
In the next step, we apply our cooling states to a multi-particle system composed of $\Nscat=10$ identical polygons. 
For this subsection only, we will assume no friction from the environment (i.e., zero damping rate), which is motivated by the difficulty of precisely modeling viscous friction and dissipation for elements with non-trivial geometrical shapes.
Figure \ref{fig:rot}(a) shows the performance of our cooling procedure applied to 10 triangles (blue), 10 squares (orange) and 10 pentagons (green), respectively. 
Here, the total energy of the system, $E_{\tot}$, reduces to its kinetic part. 
The dotted (solid) lines show the evolution of the translational (rotational) kinetic energy, $E_{\kin}^{\trans}$ ($E_{\kin}^{\rot}$) throughout the entire cooling procedure. 
This plot nicely illustrates that, at each time step, the optimal cooling state imposes the largest decrease of energy in a ``greedy'' way: the degrees of freedom that are the easiest to access get cooled first, while those that are harder to work on get cooled when it is economical to do so. 

For instance, in the case of triangular shapes, on which strong optical torques can be applied owing to their protruding edges, the procedure addresses almost simultaneously $E_{\kin}^{\trans}$ and $E_{\kin}^{\rot}$. 
In the case of squares, on which torques are harder to apply, the procedure cools first the translational degrees of freedom. 
The rotational ones are acted upon when $E_{\kin}^{\trans}$ becomes small such that cooling $E_{\kin}^{\rot}$ becomes valuable (around $t\approx 300$~$\mu$s). 
This trade-off can be observed in the bottom of Fig.~\ref{fig:rot}(b), which displays $\corr_w^{\trans}$ (blue) and $\corr_w^{\rot}$ (orange) throughout the cooling of the 10 squares (see inset). 
The procedure cools the translations during the first 300~$\mu$s and then switches to mainly cooling the rotations by applying torques, before ultimately settling into a final state around $t \approx 1$ {m}s. 
For pentagons, the applied torques are even smaller and the final state is thus only reached after $5$ {m}s (not shown).
As shown in the top of Fig.~\ref{fig:rot}(b), optimal cooling wavefronts generate complex light fields that perform a non-trivial transfer of momentum to the different particles. 
Moreover, such light fields can also create complex interactions between particles (e.g., optical binding \cite{RevModPhys.82.1767}) that are intrinsically included in our formalism and serve to cool the system down.  

\subsection{Thermal bath influence}
We now investigate the performance of our procedure in the presence of Brownian motion and damping, both resulting from a coupling to the environment [see Eq.~(\ref{eq:langevin_translation})]. 
Figure \ref{fig:rot}(c) shows the corresponding results for $\Nscat=10$ circular particles. 
To emphasize the action of optimal cooling states as compared to trivial wavefronts and to exemplify the stochastic dynamics of the nano-objects, we study the system in four successive and distinct scenarios labeled as (i) to (iv): 
(i) Initially, the nano-particles start out from a Boltzmann distribution, corresponding to a temperature $\tempenv/100$ that is hundred times smaller than the environment at $\tempenv$. 
During the first $2\damping^{-1} = 2$~ms ($t\in[0,2]$~ms, grey), the particles thermalize to $\tempenv$, i.e., to an energy given by the equipartition theorem (blue dotted line in Fig.~\ref{fig:rot}(c)). 
(ii) In the next $\approx 1$~ms, a time-constant light field of strength $P=20$~$\mu$W is applied from both ends of the waveguide along its principal transverse mode. 
As expected for such an unspecific external excitation that ``compresses'' the particles from the outside, the energy of the system is observed to increase and to settle at a high-energy steady state. 
(iii) As soon as this constant light field is turned off, the system thermalizes again and takes about $2$~ms to settle back to $\tempenv$. 
(iv) Finally, our cooling procedure is applied. 
Specifically, cooling is performed three times using three different powers of the incoming laser light: $20$~$\mu$W (orange), $2$~$\mu$W (green), and $200 $~nW (red). 
We observe the systematic cooling of the many-body system at a rate that increases the more optical power is used for cooling. 
In these cooled configurations, the velocities of the nano-particles follow a thermal distribution.

For such freely-moving many-body systems, the optical power turns out to be crucial to counteract the influence of gravity. 
When the cooling power is too weak (below $\approx 10$~nW), the particles will typically fall down as optical forces cannot compensate gravity.
Conversely, when the optical power is too strong (above $\approx 20$~$\mu$W), we sometimes observe a spatial drift of the system after it has been cooled down. 
Yet, the amplitude of this drift is very small and will impact our system on time scales that are considerably longer than typical cooling times.

\begin{figure*}[!t]
	\begin{tabular}{l{0.01\textwidth} f{0.48\textwidth} l{0.01\textwidth} f{0.48\textwidth}}
	(a)&%
		\vspace*{0pt}\hspace*{30pt}\includegraphics[scale=0.34,valign=t]{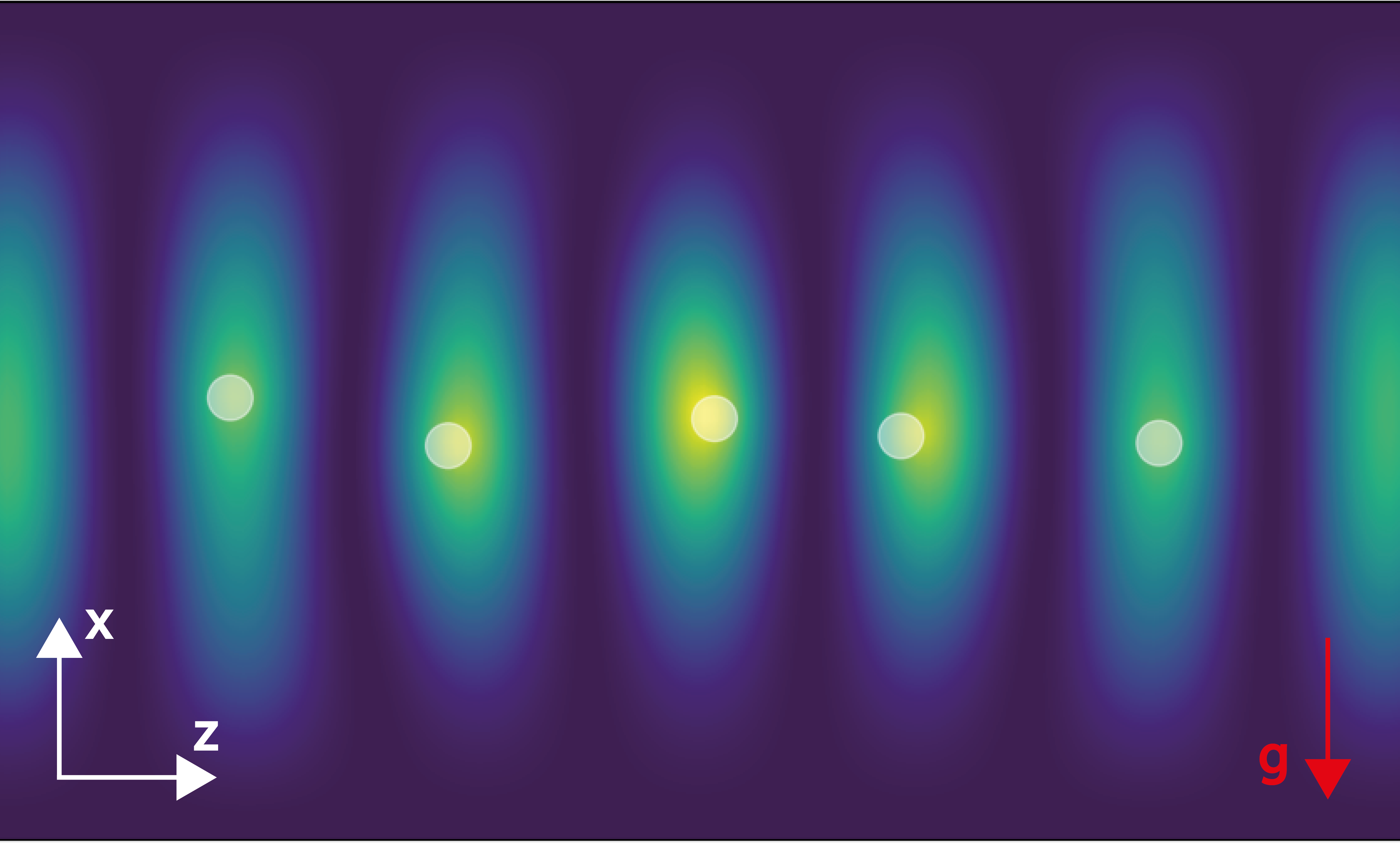}&%
    (b)&%
		\includegraphics[scale=1,valign=t]{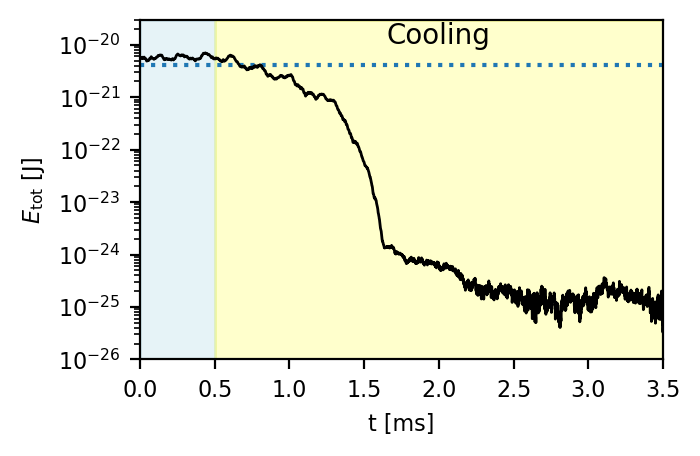}%
		\tabularnewline
	(c)&%
		\vspace*{-23pt}\includegraphics[scale=1,valign=t]{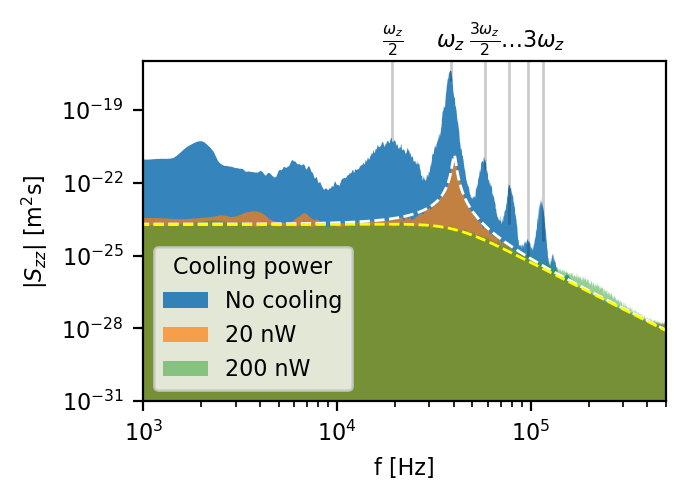}&%
    (d)&%
		\includegraphics[scale=1,valign=t]{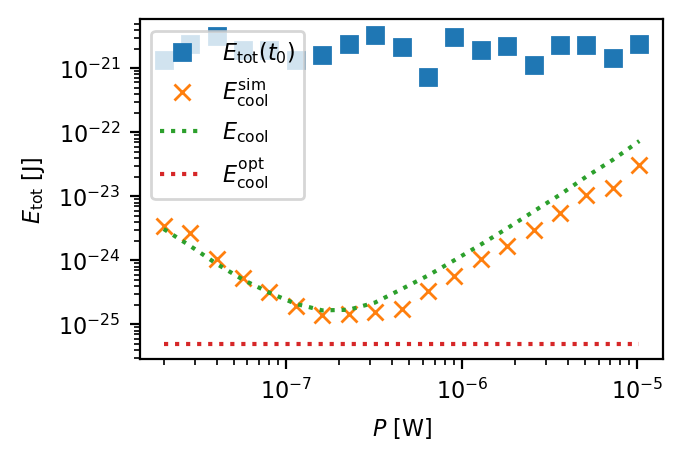}%
	\end{tabular}
	\caption{(a) $\Nscat=5$ circular nano-beads ($r=75 $~nm, white circles) are trapped by the standing-wave pattern of a laser (blue-to-yellow surface). 
	This trapping laser ($\lambda_\text{trap}=1.5 $~$\mu$m, $P_\text{trap} = 200 $~$\mu$W) gets injected from both sides of the waveguide along its fundamental transverse mode. 
	The stochastic motion of the beads is described by Eq.~(\ref{eq:langevin_translation}) ($\gamma = 6$~Hz, $T_{\mathrm{env}}=30$~K) with the additional influence of gravity ($g$, red arrow) and of the trapping field. 
	The cooling light field (not shown here) is characterized by a wavelength $\lambda = 532 $~nm and $M=10$ transverse modes.
	(b) Evolution of $E_{\tot}$ throughout the cooling process.
	In the first $500$~$\mu$s (light blue region), no cooling procedure is applied and the beads stay thermalized at $\tempenv$ (blue dashed line). 
	The procedure is launched at $t=500$~$\mu$s with a cooling power of 200~nW (yellow region) and is observed to lower $E_{\tot}$ down to $10^{-25} J$ in $\approx$ 2 ms.
	(c) Power spectral density, $|S_{zz}|$, of the spatial component $z$ for the leftmost of the 5 particles in (a) (similar behaviors are found for all the particles).
	The blue curve indicates the spectrum measured before cooling, which is characterized by a main resonance at $\omega_z$ = 40~kHz and extra harmonics (e.g., $\omega_z/2$, $3/2\omega_z$, $2\omega_z$).
	The orange (green) curve displays $|S_{zz}|$ obtained after cooling the system with a cooling power of $P=20 $~nW ($200 $~nW), while the white (yellow) dashed line describes its fitting by a Lorentzian model that is used to extract the system's temperature.    
	(d) The procedure of (a) and (b) is systematically applied to different sets of $\Nscat=5$ trapped nano-beads, while varying the cooling power $P$. 
	The total energy at the start (end) of the simulation $E_{\tot}(t_0)$ ($\Ecool^\text{sim}$) is marked by blue squares (orange crosses). The dashed green (dashed red) line depicts the estimator of the cooled energy $\Ecool$ of Eq.~(\ref{eq:endenergy1}) (the optimal cooled energy $\Ecool^{\opt}$ of Eq.~(\ref{eq:optimaltemp} for $\meancorr_w = -1$).} 
	\label{fig:trap}
\end{figure*}

\subsection{Multiple trapped particles}
\label{sec:trapped}
To illustrate the versatility of our approach, we now apply it to multiple nano-particles trapped within an optical field (rather than to freely-moving particles as above).
As shown in Fig.~\ref{fig:trap}(a), 5 nano-beads subject to Brownian motion are trapped to form a chain of opto-mechanical levitated resonators \cite{10.1088/13616633/ab6100,PhysRevLett.109.103603}. 
Specifically, a second laser, which is more intense (with $0.2 $~mW power) and which operates at a longer wavelength (with $\lambda_{\text{trap}} = 1.5$~$\mu$m) than the cooling field, gets injected simultaneously on both sides of the waveguide along its principal transverse mode. 
This field forms a standing wave, whose intensity maxima create local trapping potentials where nano-beads get confined. 
Each particle within its own trap constitutes an opto-mechanical resonator that oscillates with a frequency close to $40$~kHz along $z$ and $8.5$~kHz along $x$. 
Taken together, these nano-beads form a chain of coupled resonators, where the motion of one element affects the others \cite{liu2020prethermalization}.

Initially, the nano-particles' positions and velocities start out thermally distributed inside the trap and Fig.~\ref{fig:trap}(b) displays the time evolution of the system's total energy. 
In the first 500~$\mu$s, the system is thermalized at a level given by the equipartition theorem (dotted blue line). 
At $t=500$~$\mu$s, the procedure is applied with a power of $200$~nW and we observe the cooling of the 5 objects corresponding to a reduction of more than four orders of magnitude in their total energy in $\approx 1$~ms. 
As expected for levitated oscillators \cite{nphys1952}, this cooling approach also broadens the spectral response of each mechanical resonator along the chain. 
For instance, Fig.~\ref{fig:trap}(c) shows the power spectral density, $|S_{zz}|$, associated to the longitudinal $z$-motion of one of the five trapped particles in the absence of cooling (blue curve), which displays a main resonance at 40 kHz and where the presence of harmonics (e.g., $\omega_z/2,\dots,3\omega_z$) is indicative of the coupling amongst elements.
The orange and green curves represent the power spectral density recorded after cooling with a power of 20~nW and 200~nW, respectively. 
We observe that when the cooling power is increased, the harmonics progressively disappear and the main resonance spectrally broadens owing to the reduction of the system's centre-of-mass temperature.
Specifically, through a fitting of the power spectral densities (dashed lines in Fig.~\ref{fig:trap}(c), see appendix \ref{app:implementation}), we estimate that starting from a temperature of $\tempenv = 30$~K, the system is cooled down to effective temperatures of 23 mK and 0.7 mK for 20 nW and 200 nW, respectively.
When the optical power is increased above 200 nW, the quality factor of the resonance increases back and cooling progressively degrades (not shown).
Thus, as explained in section \ref{sec:endtemp}, we observe that cooling appears to be optimal for a power close to 200 nW.

\subsection{Cooling efficiency} 
\label{sec:endtemp}
In this section, we investigate whether the total energy reached through our procedure can be estimated based on purely theoretical considerations.
As shown in appendix \ref{sec:app:efficiency}, estimates can be derived for trapped and freely-moving particles alike. 
Under the assumption of weak damping and fast enough sampling, the average total energy obtained after the cooling process (referred to as the cooled energy) reads,
\begin{equation}
\label{eq:endenergy1}
\Ecool = C \left(\frac{ \norm[ \forcevec_{\cool}][w]^2 \coolstep + \Ndof \kB \tempenv \dofmean{\damping} }{2 \norm[ \forcevec_{\cool}][w] \meancorr_w}\right)^2\,,
\end{equation}
where $C=2$ for trapped particles and $C=1$ for free particles.
Here, $\norm[ \forcevec_{\cool}][w]^2 = \timemean{\sum_i \forcevec_{\cool,i}^2 (2 m_i)^{-1}}$ stands for the time-averaged weighted total optical force exerted on the particles and $\meancorr_w$ for the time-averaged weighted force-velocity correlation fulfilling,
\begin{equation}
\label{eq:meancorr1}
\meancorr_w = \meancorr_w^{\text{trans}} \coloneqq \frac{\timemean{\sum_i \momvec_i \cdot \forcevec_{\cool,i} (2 \mass_i)^{-1}}}{\norm[ \momvec][w]\norm[ \forcevec_{\cool}][w]}\,.
\end{equation}
In both Eqs.~(\ref{eq:endenergy1}) and (\ref{eq:meancorr1}), the brackets $\timemean{\cdot}$ indicate a time average over the physical quantities after arrival in the cooled steady state, while $\dofmean{\cdot}$ corresponds to the mean over all $\Ndof$ degrees of freedom ($\Ndof \propto \Nscat$ for identical particles).

Interestingly, the numerator of Eq.(\ref{eq:endenergy1}) is composed of two terms with different physical origins. 
The term on the left, $\norm[ \forcevec_{\cool}][w]^2 \coolstep$, shows that the cooled energy is limited by the transfer of momentum performed by the optical field to the particles at each cooling step. 
The term on the right, $\Ndof \kB \tempenv \dofmean{\damping}$, emphasizes the contribution of the thermal bath to the cooled energy. 
For different configurations of $\Nscat=5$ trapped particles [similar to Fig. \ref{fig:trap}(a)], Fig. \ref{fig:trap}(d) plots the cooled energy obtained through simulations for different cooling powers (orange crosses, $\Ecool^\text{sim}$) together with the energy prior to cooling (blue squares, $E_{\tot}(t_0)$). 
In very good agreement with the estimate provided in Eq.~(\ref{eq:endenergy1}) (dashed green, $\Ecool$), we observe a non-monotonic evolution with increasing power. 
Below $P\approx 200 $~nW, the evolution is inversely proportional to $P^2$, indicating that the fluctuations due to the thermal bath dominate (i.e., the second term of Eq.~(\ref{eq:endenergy1})). 
On the contrary, above $P\approx 200 $~nW, a quadratic ($P^2$) relation between the power and the cooled energy is observed, indicating that the transfer of momentum by the cooling field dominates (i.e., the first term of Eq.~(\ref{eq:endenergy1})).

In line with our previous observation in Fig.\ref{fig:trap}(c), we explain in appendix \ref{sec:app:endenergy} that an optimal cooling power can be identified and reveals to be function of parameters like the number of particles, the damping rate etc. 
For the system at hand, this optimal value of the cooling laser power is located near 200 nW. 
An estimator for the minimal cooled energy achievable can be obtained by optimizing Eq.~(\ref{eq:endenergy1}) with respect to $\norm[ \forcevec_{\cool}][w]$, which reads
\begin{equation}
E_{\text{cool}}^{\text{\opt}} = C (\meancorr_w)^{-2} \coolstep \dofmean{\damping} \Ndof \kB \tempenv\,,
\label{eq:optimaltemp}
\end{equation} 
where, again, $C=2$ for trapped and $C=1$ for free particles.
Assuming a perfect correlation $\meancorr_w=-1$, the optimal cooled energy of Eq.~(\ref{eq:optimaltemp}) is shown in Fig.~\ref{fig:trap}(d) as a red dashed line, which nicely predicts a lower bound of the energy obtained through simulations.
Naturally, lower energies will be achieved through faster sampling rates, $\coolstep$, or weaker couplings to the thermal bath, $\dofmean{\damping}\,$. 
Yet, amazingly, only the correlation $\meancorr_w$ depends on the parameters of the cooling procedure (like the number of optical modes and the optical wavelength). 
We thus find that the value of $\meancorr_w$ alone characterizes the performance of the routine and how to maximize it. 
Interestingly, we show in appendix \ref{sec:app:endenergy} that in the configuration where the momentum transfer of the cooling field dominates in Eq.~(\ref{eq:endenergy1}), $\Ecool$ evolves quadratically with the sampling time $\coolstep$, while $E_{\text{cool}}^{\opt}$ evolves linearly. 
Furthermore, for a fixed cooling power, there exists a threshold value $\coolstep$ below which $\Ecool$ does not improve anymore. 
A similar evolution with respect to the damping rate $\gamma$ is also presented in appendix \ref{sec:app:endenergy}.

\section{Discussion}
\label{sec:discussion}
In order to connect our theoretical results to a possible experimental realization, we discuss in the following several questions that will be relevant to such an implementation.
\subsection{Degrees of freedom}
\label{sec:scaling}
\begin{figure*}
	\begin{tabular}{l{0.01\textwidth} f{0.48\textwidth} l{0.01\textwidth} f{0.48\textwidth}}
	(a)&%
		\includegraphics[scale=1,valign=t]{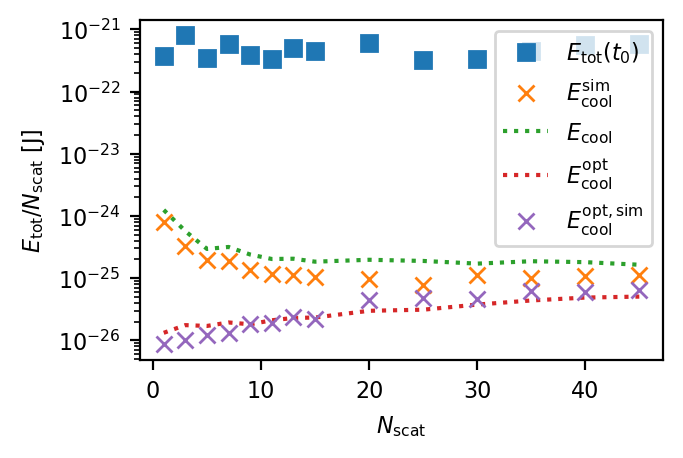}&%
    (b)&%
		\includegraphics[scale=1,valign=t]{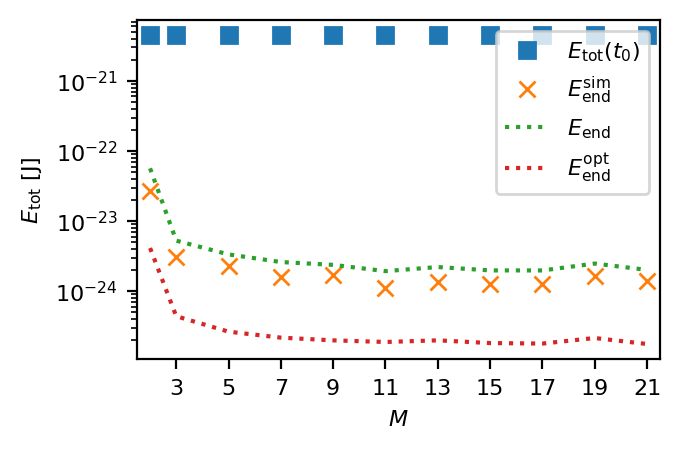}%
		\tabularnewline
	(c)&%
		\includegraphics[scale=1,valign=t]{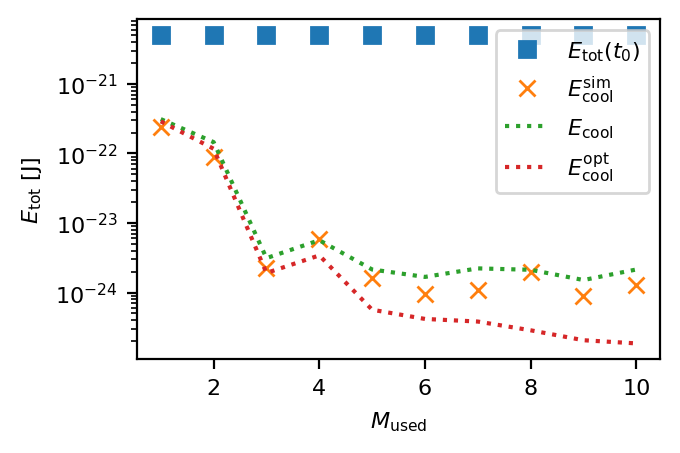}&%
    (d)&%
    \hspace*{7pt}
		\includegraphics[scale=1,valign=t]{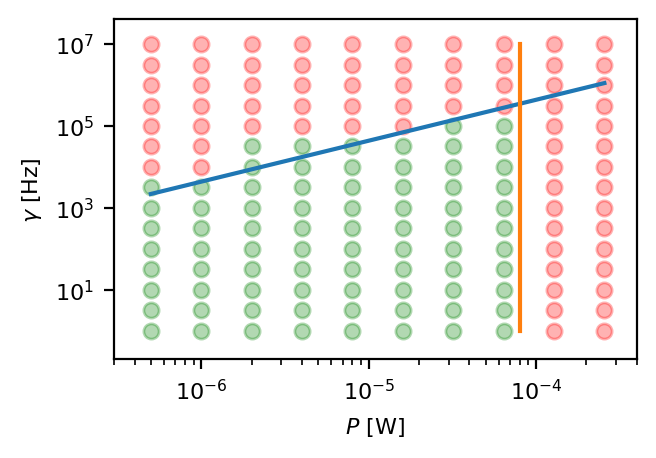}%
	\end{tabular}
\caption{(a)-(c) Total energies at the outset of the cooling procedure ($E_{\tot}(t_0)$, blue squares) and at its end ($\Ecool^\text{sim}$, orange crosses), resulting from the simulations of the systematic cooling of random initial configurations of circular nano-beads ($r=75 $~nm). 
Each configuration is averaged over ten random initial configurations.
The cooling light field is characterized by a power $P= 2 $~$\mu$W and a wavelength $\lambda = 532 $~nm. 
The dashed green (respectively dashed red) line depicts the estimator of the cooled energy $\Ecool$ of Eq.~(\ref{eq:endenergy1}) (respectively the optimal cooled energy $\Ecool^{\opt}$ of Eq.~(\ref{eq:optimaltemp})).
In (a), the number of particles, $\Nscat$, is varied while the number of propagating modes $\modes=10$ is fixed and the purple crosses ($\Ecool^{\opt,\text{sim}}$) report cooling performed under the optimal conditions of Eq.~(\ref{eq:optimaltemp}).
In (b), $\modes$ is varied while we set $\Nscat=10$. 
In (c), $\Nscat=10$ and $\modes=10$, while we vary the number of modes  $\modes_{\text{used}}$ used to assemble the energy-shift operator $\Qt$.
(d) Efficiency of the cooling of $\Nscat=10$ nano-beads, while varying the damping rate, $\gamma$, and the laser power $P$. 
The green dots indicate when the cooled energy is at most half the initial energy (red dots show when this is not the case). 
The blue and orange lines indicate the bounds provided by Eq.~(\ref{eq:bounds}).}
\label{fig:numscat}
\end{figure*}
The first topic we address here regards how many optical modes are required in order to achieve cooling. 
Quite remarkably, it turns out that the degrees of freedom of the cooling wavefronts are not directly linked to the particles' degrees of freedom. 
In other words, even just a few optical modes can be harnessed to cool multiple nano-particles simultaneously (a feature that should simplify an experimental implementation). 
 
To illustrate this point, we show in Fig.~\ref{fig:numscat}(a) cooling simulations when the number of particles, $\Nscat$, is varied and the number of modes is set to $\modes=10$. 
Here, instead of $E_\tot$, we plot the ratio $E_\tot/\Nscat$ (indicative of individual particles' energies) in order to emphasize the evolution of cooling performance with the number of particles.  
The blue squares represent the averaged simulated initial energies ($E_{\tot}(t_0)$) and the orange crosses represent the averaged end energies using a cooling power of 2 $\mu$W ($\Ecool^\text{sim}$) while the dashed green line provides the corresponding estimator of Eq.~(\ref{eq:endenergy1}) ($\Ecool$). 
For each value of $\Nscat$, the cooling power is also adjusted to reach optimal cooling conditions and the purple crosses mark the simulated averaged end energies ($E_{\cool}^{\opt,\text{sim}}$), while the dashed red line provides the corresponding estimator of Eq.~\eqref{eq:optimaltemp} ($E_{\text{cool}}^{\text{opt}}$).
Remarkably, on the one hand, for a fixed power of $P=2$~$\mu$W, $\Ecool /\Nscat$ stays approximately constant, irrespectively of $\Nscat$. 
Yet, on the other hand, the optimal cooling energy per particle linearly increases with $\Nscat$.
In short: the procedure can handle more nano-particle degrees of freedom than the degrees of freedom available in the light field (i.e., $\modes$), but the optimal cooling performances one can expect will progressively degrade when more particles are added. 

We now reproduce these simulations for the case of $\Nscat=10$ nano-particles and when the width of the waveguide (and thus the number of $\modes$) is increased. 
The corresponding results in Fig.~\ref{fig:numscat}(b) emphasize that only a small number of optical modes is required to cool the system down. 
Starting from $\modes=2$ to 3, the cooled energy decreases by one order of magnitude and does not significantly improve for higher numbers of modes.

\subsection{Missing information}
\label{sec:missing}
An important aspect of our procedure is the fact that it is also very robust to a reduction of both the degree of control exerted over the input field as well as of the information collected in the scattered field. 
Such robustness is of particular importance to apply the procedure to nano-particles evolving in free space (not confined in a waveguide), where the control of the input and the collection of the output field scattered by the particles are both intrinsically limited by the numerical aperture of the optical setup. 
In order to simulate a partial control and collection of the field, part of the scattering matrix will be ignored. 
In Fig. \ref{fig:numscat}(c), we show the cooling of an initial random configuration made of $\Nscat=10$ nano-particles in a waveguide with $\modes=10$ propagating modes [color code similar to Figs.~\ref{fig:numscat}(a) and (b)]. 
The simulations are run while only making use of the scattering information of a reduced number of modes ($\modes_{\text{used}}$). 
We observe that, while cooling degrades when only 1 or 2 modes are open, it quickly converges to an energy similar to the one obtained when the complete $\S$ matrix is used. 
This emphasizes that, in order to be efficient, the procedure only relies on a small number of modes carrying sufficient information to operate.

\subsection{Cooling conditions}
At last, we focus on the operating range of our approach, i.e., the parameter range over which cooling remains efficient. 
Using the estimator of $\Ecool$ provided in Eq.~(\ref{eq:endenergy1}), we derive in appendix \ref{app:coolinglimits} different operating bounds for the relevant system parameters such as the laser power $P$ or the damping rate $\gamma$. 
To illustrate these bounds, we show in Fig.~\ref{fig:numscat}(d) a map of the efficiency of our procedure as a function of these two parameters. 
In this plot the green dots indicate when cooling is efficient in the sense that the system's energy is at least reduced by $50\%$ from its initial value (i.e., $\Ecool < \Ndof k_B \tempenv / 4$). 
The red dots indicate when this condition is not fulfilled. 
The boundary between the green and the red regions indicates the operating range of our approach. 
In appendix \ref{app:coolinglimits}, we show that estimators for these bounds can be obtained with the following relations,
\begin{equation}
	\frac{2 \meancorr_w^2}{ (\dofmean{\damping})^2 } > \frac{\Ndof \kB T}{\norm[ F_{\cool}][w]^2} > \frac{\coolstep^2}{\meancorr_w^2}\,.
	\label{eq:bounds}
\end{equation}

The (averaged) estimates of the bounds on $P$ and $\damping$ deduced from Eq.~(\ref{eq:bounds}) are marked in Fig.~\ref{fig:numscat}(d) by the orange and blue lines. 
Physically, the orange line delimits the influence of momentum transfer from the light field: when $P$ is too strong, too much momentum is transferred to the nano-particles in each cooling step and the system gains more energy than it loses. 
In turn, the blue line describes the impact of the coupling to the environment: when $\gamma$ lies above the blue line, more energy gets added by the Langevin forces (i.e., stochastic term in Eq.~(\ref{eq:langevin_translation})) than what is being removed by the cooling force. 
Even though cooling remains efficient within the operating areas, its performance degrades the closer one gets to the operating bounds (blue and orange lines in  Fig.~\ref{fig:numscat}(d)). 

Similar estimators can be obtained for all relevant parameters. 
For the sampling rate $\coolstep$, we can see from Fig.~\ref{fig:numscat}(d) that the procedure becomes inefficient for $\damping > 0.5$~MHz~$= 1/(2 \coolstep)$. 
Yet, this condition is necessary, but not sufficient. 
Indeed, between two consecutive measurements (at $t$ and $t+\coolstep$), the modification of $\S$ must remain small. 
In short, the displacement of the particles moving at mean speed $v$ should remain small compared to the optical wavelength $\lambda$, or more specifically: 
\begin{equation}
v \coolstep < \lambda/4\,.
\label{eq:condition}
\end{equation}
Translating this relation to the case, where only translational degrees of freedom contribute, leads to the following bound in sampling time (see appendix \ref{app:lambda} for details),
\begin{equation}
\label{eq:condition:start}
\coolstep < \frac{1}{4}\sqrt{\frac{\sum_i \mass_i \lambda^2}{\Ndof\kB\tempenv}}\,.
\end{equation}

\section{Conclusion}
In this work, we present a new many-body cooling scheme and explore its potential to efficiently cool dielectric particles under realistic conditions. 
In particular, we show that, through the use of the energy-shift operator, wavefront shaping can be harnessed to cool down both translational and rotational degrees of freedom on multiple levitated objects at once. 
We demonstrate that our method applies to freely moving and trapped nano-objects alike, regardless of their geometries \cite{PhysRevLett.117.123604,s41565-019-0605-9}. 
Both on the analytical level and in simulations, we are able to derive estimates of the cooling efficiency and to demonstrate the existence of optimal cooling conditions. 
Moreover, we quantify the existence of parameter ranges, outside of which our scheme becomes inefficient. 
At last, we show that our approach is remarkably robust to a partial measurement and to a partial control of the asymptotic field and that it operates efficiently on a large number of particles even when just a few optical modes are accessible. 

To the best of our knowledge, this cooling technique represents the first strategy that is able to address multiple coupled nano-objects in levitation at once. 
We thus anticipate that it opens the door to experimental progress in the field of optical levitation \cite{10.1088/13616633/ab6100}. 
For instance, our scheme could serve as a protocol to reach for the first time the motional quantum ground state of a many-body system and thus realize the entanglement of two or more levitated objects \cite{science.abg3027}. 

\section{Acknowledgments}
We thank J.~Bertolotti and Y.~Louyer for helpful discussions and the team behind the open-source code NGSolve for assistance. Support by the Austrian Science Fund (FWF) under Project No.~P32300 (WAVELAND) and funding for NB from the European Union’s Horizon 2020 research and innovation program under the Marie Skłodowska-Curie grant agreement No.~840745 (ONTOP) are gratefully acknowledged. The computational results presented were achieved using the Vienna Scientific Cluster (VSC).
%\clearpage
\bibliographystyle{apsrev4-2}
\bibliography{references} 
\clearpage

%%%%%%%%%%%%%%%%%%%%%%%%%%%%%%%%%%%%%%%%%%%%%%%%%%%%%%%%%%%
%%%%%%%%%%%%%%%%%%%%%%%%%%%%%%%%%%%%%%%%%%%%%%%%%%%%%%%%%%%
%%%%%%%%%%%%%%%%%%%%%%%%%%%%%%%%%%%%%%%%%%%%%%%%%%%%%%%%%%%
%%%%%%%%%%%%%%%%%%%%%%%%%%%%%%%%%%%%%%%%%%%%%%%%%%%%%%%%%%%
%%%%%%%%%%%%%%%%%%%%%%%%%%%%%%%%%%%%%%%%%%%%%%%%%%%%%%%%%%%
%%%%%%%%%%%%%%%%%%%%%%%%%%%%%%%%%%%%%%%%%%%%%%%%%%%%%%%%%%%
%%%%%%%%%%%%%%%%%%%%%%%%%%%%%%%%%%%%%%%%%%%%%%%%%%%%%%%%%%%
%%%%%%%%%%%%%%%%%%%%%%%%%%%%%%%%%%%%%%%%%%%%%%%%%%%%%%%%%%%
%%%%%%%%%%%%%%%%%%%%%%%%%%%%%%%%%%%%%%%%%%%%%%%%%%%%%%%%%%%
%%%%%%%%%%%%%%%%%%%%%%%%%%%%%%%%%%%%%%%%%%%%%%%%%%%%%%%%%%%
%%%%%%%%%%%%%%%%%%%%%%%%%%%%%%%%%%%%%%%%%%%%%%%%%%%%%%%%%%%
%%%%%%%%%%%%%%%%%%%%%%%%%%%%%%%%%%%%%%%%%%%%%%%%%%%%%%%%%%%
\appendix

\section{Energy-shift relation}
\label{app:GWSrelation}
Within a given domain $\Omega$ of boundary $\partial \Omega$, we denote by $(\E_i,\H_i)$ the solution of Maxwell's equations provided in Eq.~(\ref{eq:PDE:Efield}) for a given permittivity $\epsilon_i$ and permeability $\mu_i$. 
For two distinct complex amplitudes $(\E_i,\H_i)_{i\in\{1,2\}}$, corresponding to two distinct configurations $(\epsilon_i,\mu_i)_{i\in\{1,2\}}$, we define the Hermitian form over $\Omega$ as
\begin{equation}
\label{eq:innerdef}
\inner[(\E_1,\H_1)][(\E_2,\H_2)][\partial \Omega] \coloneqq \frac{1}{2} \int_{\partial \Omega} (\E_1^* \times \H_2 - \H_1^* \times \E_2) \cdot \d \n,
\end{equation}
in which $\n$ depicts the outgoing normal to the surface $\partial \Omega$. 
If both fields are equal, then Eq.~(\ref{eq:innerdef}) corresponds to the integrated time-averaged Poynting vector and disappears for systems without loss or gain. 
Combining Eq.~(\ref{eq:PDE:Efield}) with the relation $\nablavec \cdot (\vec{A}\times \vec{B}) = \vec{B} \cdot (\nablavec \times \vec{A}) -  \vec{A} \cdot (\nablavec \times \vec{B})$, the Hermitian form of Eq.~(\ref{eq:innerdef}) can be recast as
\begin{equation}
\label{eq:flowfield}
\begin{split}
\inner[(\E_1,\H_1)][(\E_2,\H_2)][\partial \Omega] =& \frac{\i \omega}{2} \int_{\Omega}    (\mu_2- \mu_1^*) \H_1^\dagger \H_2 \\
&+  (\epsilon_2 - \epsilon_1^*) \E_1^\dagger \E_2 \d V.
\end{split}
\end{equation}
We now assume that $\epsilon_i$ and $\mu_i$ are real (i.e., lossless media) and functions of a parameter $\alpha$ (this parameter will later be chosen to be the time $t$). 
Moreover, if we choose the two configurations $(\epsilon_i, \mu_i)_{i \in \{1,2\}}$ to be close to one another, we can make the first-order approximation,
\begin{equation}
\label{eq:approx:media}
\begin{split}
    (\epsilon_2, \mu_2) \approx (\epsilon_1, \mu_1) + \Delta \alpha (\partial_\alpha \epsilon_1, \partial_\alpha \mu_1)\,, 
\end{split}
\end{equation}
from which we can readily deduce
\begin{equation}
\label{eq:approx:field}
\begin{split}
    (\E_2,\H_2) \approx (\E_1,\H_1) + \Delta \alpha (\partial_\alpha \E_1,\partial_\alpha\H_1)\,.
\end{split}
\end{equation}
Inserting Eqs.~(\ref{eq:approx:media}) and (\ref{eq:approx:field}) into Eq.~(\ref{eq:flowfield}), while also making use of the fact that Eq.~\eqref{eq:innerdef} is Hermitian, we deduce by comparing the coefficients of $\Delta \alpha$ that,
\begin{equation}
\label{eq:inner_int_relation}
\begin{split}
\inner[(\E,\H)][(\der[\alpha]\E,\der[\alpha]\H)][\partial \Omega] = \i \frac{\omega }{2}\int_{\Omega}& |\E|^2  \der[\alpha] \epsilon \\
+& |\H|^2  \der[\alpha] \mu, 
\end{split}
\end{equation}
in which $(\epsilon_1, \mu_1)$ and $(\E_1,\H_1)$ were relabelled $(\epsilon, \mu)$ and $(\E,\H)$, respectively, and the left-hand side is obtained by substituting $(\der[\alpha]\E,\der[\alpha]\H)$ for $(\E_2,\H_2)$ in Eq.~\eqref{eq:innerdef}.

\subsection{Energy change}
We now turn to a scattering system made of scatterers moving as a function of time $t$ and we substitute in Eq.~(\ref{eq:inner_int_relation}) the parameter $\alpha$ by the time $t$. 
For easier notation, we will now proceed with the case of dielectric scatterers. 
The time-varying medium is described by a constant permeability of $\mu=\mu_0$ and a real permittivity $\epsilon(\vec{r},t)$ that is a function of space $\vec{r}$ and time $t$. 
Since the system is made up of distinct rigid scatterers, we decompose $\epsilon(\vec{r},t)$ as
\begin{equation}
\epsilon(\vec{r},t) = \epsilon_0 + \sum_i \epsilon_{i}(\vec{r},t)\,,
\end{equation}
where $\epsilon_{i}(\vec{r},t)$ depicts the permittivity of the $i$th nano-particle relative to the background. 
The motion of the rigid particles is assumed to be slow with respect to the time scale of the light field. 
To illustrate the movement of a single particle, we sketch in Fig.~(\ref{fig:material_constants}) the evolution of $\epsilon_{i}(\vec{r},t)$ for the case of a rectangle with a center of mass $\posvec_i(t)$, whose locations at two instances of time ($t_0$ and $t$) are displayed in blue and black, respectively. 
Throughout the particle's motion, a coordinate $\vec{x}$ in the rectangle at $t_0$ will change into a coordinate, 
\begin{equation}
\label{app:eq:displacement}
    \vec{r}_{i}(\vec{x},t) =\posvec_i(t) + \rotmatrix(\angvec_i(t))[\vec{x}-\posvec_i(t_0)]\,,
\end{equation}
at time $t$, where $\rotmatrix(\angvec_i(t))$ stands for the rotation matrix for the instantaneous Euler angles $\angvec_i(t)$ of the rigid body at time $t$.
Assuming rigid bodies with time-constant properties, the material derivative fulfills:
\begin{equation}
\label{eq:materialderivative}
0 = \frac{D \epsilon_i}{D t} = \der[t] \epsilon_i + (\der[t] \vec{r}_{i}) \cdot \nablavec \epsilon_i\,.
\end{equation}
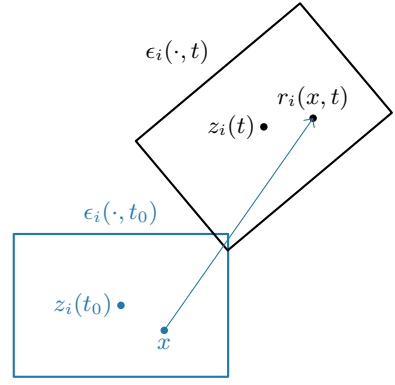
\begin{figure}
    \centering
    \begin{tikzpicture}[line cap=round,line join=round,>=triangle 45,x=0.95cm,y=0.95cm]
        \clip(0,0) rectangle (6,6);	
        \pgfmathsetmacro{\xlength}{6}		
        \pgfmathsetmacro{\ylength}{6}		
        
        \pgfmathsetmacro{\squarelength}{6}			
        \pgfmathsetmacro{\marginx}{(\xlength - \squarelength)/2 }
        \pgfmathsetmacro{\marginy}{(\ylength-\squarelength)/2}	
        \pgfmathsetmacro{\originty}{\ylength/2}	
        \pgfmathsetmacro{\origintxA}{\marginx+\squarelength/2}
        %%%%%%%%%%%%%%rectangle dim%%%%%%%%%%%
        \pgfmathsetmacro{\lenx}{3}
        \pgfmathsetmacro{\leny}{2}
        %%%%%%%%%%%%%%%orientation time%%%%%%%%%%%%%%%%%%%
        \pgfmathsetmacro{\zxA}{-1}
        \pgfmathsetmacro{\zyA}{-1.5}	
        \pgfmathsetmacro{\rotateA}{0}
        \pgfmathsetmacro{\zxB}{1}
        \pgfmathsetmacro{\zyB}{1}	
        \pgfmathsetmacro{\rotateB}{40}
        %%%%%%%%%%%%%%position point%%%%%%%%%%%
        \pgfmathsetmacro{\prad}{0.7}
        \pgfmathsetmacro{\pang}{-30}	
        %%%%%%%%%%%%%t_0%%%%%%%%%%
        \pgfmathsetmacro{\origintx}{\origintxA}
        \pgfmathsetmacro{\zx}{\zxA}
        \pgfmathsetmacro{\zy}{\zyA}	
        \pgfmathsetmacro{\rotate}{\rotateA}

        \definecolor{mplgreen}{HTML}{2ca02c}
        \definecolor{mplblue}{HTML}{1f77b4}
        \definecolor{mplorange}{HTML}{ff7f0e}
        
        \draw[mplblue,thick,rotate around={\rotate:({\origintx+\zx},{\originty+\zy})}] ({\origintx+\zx-\lenx/2},{\originty+\zy-\leny/2}) rectangle (({\origintx+\zx+\lenx/2},{\originty+\zy+\leny/2});		
        \node[mplblue,circle,fill,inner sep=1pt] at ({\origintx+\zx},{\originty+\zy}){};		
        \node[mplblue,anchor=east] at ({\origintx+\zx},{\originty+\zy}){$z_i(t_0)$};		
        \node[mplblue,circle,fill,inner sep=1pt] at ({\origintx+\zx+\prad*cos(\rotate+\pang)},{\originty+\zy+\prad*sin(\rotate+\pang)}){};	
        \node[mplblue,anchor=north] at ({\origintx+\zx+\prad*cos(\rotate+\pang)},{\originty+\zy+\prad*sin(\rotate+\pang)}){$x$};	
        \node[mplblue,anchor=south] at ({\origintx+\zx+\leny/2*cos(\rotate+90)},{\originty+\zy+\leny/2*sin(\rotate+90)}){$\epsilon_i(\cdot,t_0)$};	
        %%%%%%%%%%%%%t_1%%%%%%%%%%
        \pgfmathsetmacro{\origintx}{\origintxA}
        \pgfmathsetmacro{\zx}{\zxB}
        \pgfmathsetmacro{\zy}{\zyB}	
        \pgfmathsetmacro{\rotate}{\rotateB}
        
        \draw[black,thick,rotate around={\rotate:({\origintx+\zx},{\originty+\zy})}] ({\origintx+\zx-\lenx/2},{\originty+\zy-\leny/2}) rectangle (({\origintx+\zx+\lenx/2},{\originty+\zy+\leny/2});		
        \node[black,circle,fill,inner sep=1pt] at ({\origintx+\zx},{\originty+\zy}){};		
        \node[black,anchor=east] at ({\origintx+\zx},{\originty+\zy}){$z_i(t)$};		
        \node[black,circle,fill,inner sep=1pt] at ({\origintx+\zx+\prad*cos(\rotate+\pang)},{\originty+\zy+\prad*sin(\rotate+\pang)}){};	
        \node[black,anchor=south] at ({\origintx+\zx+\prad*cos(\rotate+\pang)},{\originty+\zy+\prad*sin(\rotate+\pang)}){$r_i(x,t)$};
        \node[black,anchor=south east] at ({\origintx+\zx+\leny/2*cos(\rotate+90)},{\originty+\zy+\leny/2*sin(\rotate+90)}){$\epsilon_i(\cdot,t)$};
        %%%%%%%%%%%%Arrow%%%%%%%%%%%%%%%%%%
        \draw[mplblue,arrows={- angle 90}] ({\origintxA+\zxA+\prad*cos(\rotateA+\pang)},{\originty+\zyA+\prad*sin(\rotateA+\pang)}) -- ({\origintx+\zxB+\prad*cos(\rotateB+\pang)},{\originty+\zyB+\prad*sin(\rotateB+\pang)});
    \end{tikzpicture}
    \caption{The blue shape depicts a rigid dielectric of rectangular shape at time $t=t_0$, which is tagged with a permittivity $\epsilon_i(.,t_0)$ and whose center of mass is labelled $z_i(t_0)$. 
    The black shape depicts the position of the particle at a time $t$ (permittivity $\epsilon_i(.,t)$ and center of mass $z_i(t)$). 
    Throughout the motion of the particle, a location $x$ inside the dielectric at $t=t_0$ gets displaced to $r_i(x,t)$ at time $t$.}
    \label{fig:material_constants}
\end{figure}
We denote by $\matrix{\moi}_{i}(\angvec_i(t))$ and $\angmomvec_i(t)$ the matrix of inertia and the angular momentum of the object, respectively. 
Using the relation $\der[t] (\rotmatrix(\angvec_i(t))) = (\matrix{\moi}_i^{-1}\angmomvec_i)(t)  \times \rotmatrix(\angvec_i(t))$ (see Ref.~\cite{martinetz2018gas}), we derive by using Eq.~\eqref{app:eq:displacement} that
\begin{equation}
\label{eq:energychange:posderivative}
\der[t]\vec{r}_i = \mass_i^{-1} \momvec_i + (\matrix{\moi}_i^{-1}\angmomvec_i)  \times (\vec{r}_i-\posvec_i)\,.
\end{equation}
Combining the identity $\vec{A} \cdot (\vec{B} \times \vec{C}) = \vec{B} \cdot (\vec{C} \times \vec{A})$ with Eqs.~(\ref{eq:materialderivative}) and (\ref{eq:energychange:posderivative}), we then obtain
\begin{equation}
\label{eq:eps_derivative}
\begin{split}
\der[t]\epsilon =& - \sum_i \matrix{\moi}_i^{-1} \angmomvec_i  \cdot [(\vec{r}_{i}-\posvec_i) \times \nablavec \epsilon] \\
&- \sum_i \mass_i^{-1} \momvec_i \cdot \nablavec \epsilon\,.
\end{split}
\end{equation}
We emphasize here that $\vec{r}_{i}-\posvec_i$ corresponds to the lever used for calculating the torque of the particles.
A similar relation can also be derived for non-constant $\mu$. 
The mean force density exerted on an incompressible particle by a monochromatic wave reads \cite[p. 242]{landau2013electrodynamics},  
\begin{equation}
\label{eq:forcedensity}
\vec{f} =  - \frac{1}{4} \left( |\E|^2 \nablavec\epsilon +   |\H|^2 \nablavec \mu\right)\,.
\end{equation}
Combining Eqs.~(\ref{eq:eps_derivative}) and (\ref{eq:forcedensity}), the relationship of Eq.~(\ref{eq:inner_int_relation}) can be recast as,
\begin{equation}
\begin{split}
\frac{\inner[(\E,\H)][(\der[t]\E,\der[t]\H)][\partial \Omega]}{2 \omega \i} =& \sum_i \mass_i^{-1}  \forcevec_{\cool,i} \cdot \momvec_i   \\
&+ \sum_i \torquecool[i] \cdot (\matrix{\moi}_i^{-1} \angmomvec_i)\,,
\label{eq:demenenrgyrelation}
\end{split}
\end{equation}
in which $\forcevec_{\cool,i}$ and $\torquecool[i]$ describe the optical force and torque applied by the cooling field on the $i$th particle, respectively. 
Furthermore the right-hand side of Eq.~\eqref{eq:demenenrgyrelation} can be related to the power transferred to the particles through the cooling field. 
From Newton's second law, the total energy (defined as $E_{\mathrm{tot}} = E_{\kin} + E_{\pot}$) relates to the  power of non-conservative forces through the relation,
\begin{equation}
    \der[t] E_{\mathrm{tot}} = \sum_i \frac{\forcevec_{\cool,i} \cdot \momvec_i}{\mass_i} + \sum_i \torquecool[i] \cdot (\matrix{\moi}_i^{-1} \angmomvec_i) + P_{\text{\text{nc}}}\,.
    \label{eq:etot}
\end{equation}
In Eq.~(\ref{eq:etot}), $P_{\text{\text{nc}}}$ stands for the power of all non-conservative forces other than the cooling force acting on the particles. 
Injecting Eq.~(\ref{eq:etot}) into Eq.~(\ref{eq:demenenrgyrelation}), we get the energy-shift relation 
\begin{equation}
\label{eq:inner_energy}
\begin{split}
-\i\inner[(\E,\H)][(\der[t]\E,\der[t]\H)][\partial \Omega] =& \,2 \omega (\der[t] E_\tot - P_{\text{nc}}).
\end{split}
\end{equation}

\subsubsection{Waveguide confinement}
The energy-shift relation of Eq.~(\ref{eq:inner_energy}) is now applied to the case of dielectric scatterers moving inside a hollow waveguide \cite[p. 549]{pollack}, to which two semi-infinite leads are attached on its left and right ends. 
The field propagates along the $z$-direction.
The lateral boundaries are made up of a perfect conductor at which the tangential components of the electric field and the normal components of the magnetic field disappear. 
We denote by $\Gamma$ the cross-section of a hollow waveguide lead connected to the far field and by $\n$ the normal vector of the surface pointing in the outgoing direction. 
The waveguide supports a finite number of propagating modes, which depends on the wavelength. The waves can be split up into Transverse Magnetic (TM) and Transverse Electric (TE) parts, where the magnetic and electric fields are orthogonal to the direction of propagation, respectively. 
Here, the product $\inner[(\E_1,\H_1)][(\E_2,\H_2)][\Gamma]$ given in Eq.~\eqref{eq:innerdef} is obtained by replacing $\partial \Omega$ with $\Gamma$ and fulfills
\begin{equation}
\begin{split}
	&\inner[(\E_1,\H_1)][(\E_2,\H_2)][\Gamma] = \\
	&\frac{\i}{2 \omega} \int_{\Gamma} \mu_0^{-1} \E_{1}^{\text{TE},\dagger} \der[z]\E_{2}^\text{TE} - \mu_0^{-1}  \der[z] \E_{1}^{\text{TE},\dagger} \E_{2}^\text{TE} \\
	&+ \epsilon_0^{-1} \H_{1}^{\text{TM},\dagger} \der[z]\H_{2}^\text{TM} - \epsilon_0^{-1} \der[z] \H_{1}^{\text{TM},\dagger} \H_{2}^\text{TM} \d \sigma\,.
	\label{eq:TETM}
\end{split}
\end{equation}
We now consider $\text{TE}$ waves, 
\begin{equation}
\begin{split}
    \E^\text{TE} =& \sum_i \vec{\eta}_i(x,y)(c_{i}^{\text{TE},\text{in}}e^{\i k_{z,i} z} + c_{i}^{\text{TE},\text{out}}e^{-\i k_{z,i} z}),
    \label{eq:TETM_2}
\end{split}
\end{equation}
with $\int_{\Gamma} \vec{\eta}_i^{\,\dagger} \vec{\eta}_j = \frac{\omega \mu_0}{k_{z,i}} \delta_{i,j}$ and $\vec{\eta}_i \perp \n$. 
This normalization sets the ``Ponyting vector'' of each basis function, integrated over the lead diameter, equal to $1$. 
With this, we can see that the Hermitian form of Eq.~\eqref{eq:TETM} is 0 if $(\E_1,\H_1)$ is incoming while $(\E_2,\H_2)$ is outgoing (or the other way around). 
An equation similar to Eq.~(\ref{eq:TETM_2}) can be readily derived for TM waves. 
Thus, by combining Eq.~\eqref{eq:TETM} for all leads, we can write,
\begin{equation}
	\inner[(\E_1,\H_1)][(\E_2,\H_2)][\partial \Omega] = \c_{1}^{\,\text{out},\dagger} \c_{2}^{\,\text{out}} -\c_{1}^{\,\text{in},\dagger} \c_{2}^{\,\text{in}}\,,
	\label{eq:enenrgyshift2}
\end{equation}
in which the vectors $\c_{1/2}^{\,\text{in/out}} $ encapsulate both the field coefficients introduced in Eq.~(\ref{eq:TETM_2}) for the TE parts and similar ones for the TM parts of the wave for all leads. 
Specifically, $\c_{1/2}^{\,\text{in}}$ denotes the incoming wavefronts that are injected into the waveguide, while $\c_{1/2}^{\,\text{out}}$ labels the outgoing wavefronts that are scattered away. 
Any couple of incoming/outgoing wavefronts is connected by the scattering matrix through the relation $\c^{\,\text{out}} = \S \c^{\,\text{in}}$. 
Thus, Eq.~(\ref{eq:enenrgyshift2}) can be recast as, 
\begin{equation}
\label{eq:inner_Smatrix}
\inner[(\E_1,\H_1)][(\E_2,\H_2)][\partial \Omega] = \c_{1}^{\,\text{in},\dagger} (\S_1^\dagger \S_2 - \eye) \c_{2}^{\,\text{in}}\,,
\end{equation}
in which $\S_1$ and $\S_2$ correspond to two distinct configurations $(\epsilon_i, \mu_i)_{i \in \{1,2\}}$ and thus two scattering organizations, respectively. 
If these two configurations are close to one another, we note $\S_1=\S(t)$ and $\S_2 = \S(t+\Delta t) \approx \S(t) +\Delta t \partial_t \S(t)$, while setting $\c_{2}^{\,\text{in}} = \c_{1}^{\,\text{in}} = \c^{\,\text{in}}$. 
Now, combining this expression with Eqs.~\eqref{eq:inner_Smatrix} and \eqref{eq:inner_energy} and using the unitarity of $S(t)$, we get Eq.~(\ref{eq:GWSrelation}) of the main text,
\begin{equation}
\label{eq:ES_app}
	\c^{\,\text{in},\dagger} \Qt \c^{\,\text{in}} = 2 \omega (\der[t] E_\tot - P_{\text{nc}})\,.
\end{equation}

\subsubsection{Free space}
\label{sec:app:freespace}
The energy-shift relation can also be derived for particles evolving in free space (not confined in a waveguide). 
According to Ref.~\cite{saxon1955tensor}, the electrical field scattered into the far field by an object of arbitrary shape reads
\begin{equation}
\E(\n r) \sim \left(\frac{\mu_0}{\epsilon_0}\right)^{1/4} \left[\vec{F}^{\,\text{in}}(\n) \frac{e^{-\i k r}}{r} + \vec{F}^{\,\text{out}}(\n) \frac{e^{\i k r}}{r} \right]\,,
\end{equation}
in which $\vec{F}^{\,\text{in}}(\n)$ and $\vec{F}^{\,\text{out}}(\n)$ are vectors characterizing the incoming and outgoing components of the field oriented in a direction  $\vec{F}^{\,\text{in}/\text{out}}\perp \n$,  respectively. 
We define the tensor scattering matrix for the surface integral over the unit sphere (all directions of incoming waves)
\begin{equation}
\vec{F}^{\,\text{out}}(\n) = - \int_{|\n'|=1} \S(\n,\n') \vec{F}^{\,\text{in}}(-\n') \d S'\,.
\end{equation}
For moving particles, the energy-shift relation reads
\begin{equation}
\begin{split}
	\int_{|\n'|=1} \int_{|\n''|=1}& \vec{F}_{\,\text{in}}^\dagger(\n'') \Qt(\n'',\n') \vec{F}_{\,\text{in}}(\n') \d S'' \d S' \\
	=&  2 \omega (\der[t] E_\tot  - P_{\text{nc}}),
\end{split}
\label{eq:FreeSpaceGWS}
\end{equation}
in which the energy-shift operator is now defined as $\Qt(\n'',\n') = -\i\int_{|\n|=1} \S^\dagger(-\n'',\n) \der[t]\S(\n,-\n')\d S$. 
From Eq.~\ref{eq:FreeSpaceGWS}, we conclude that our cooling technique can be applied to particles experiencing 3D free-space motions (i.e., not confined in a waveguide). 

\section{Non-spherical particles}
\subsection{Stochastic motion}
\label{app:complete_description}
In section \ref{sec:model} of the main text, we described the stochastic motion of nano-particles while restricting ourselves to the case of spherical objects (i.e., without rotational degrees of freedom). 
Here, we describe the motion of arbitrary rigid particles in partial vacuum. 
On top of translational degrees of freedom, we are now describing 3D rotations. 
We consider particles that are well separated spatially and whose motions only interact through the scattering of the light field. 
Each particle has a mass $\mass_i$ and a moment of inertia $\matrix{\moi}_{i,0} = \Diag(\moi_{i,1},\moi_{i,2},\moi_{i,3})$ in a reference orientation. 
The positions $\posvec_i$ and the Euler angles $\angvec_i$ set the tensor of inertia $\matrix{\moi}_i(\angvec_i) = \rotmatrix(\angvec_i) \matrix{\moi}_{i,0} \rotmatrix^T(\angvec_i)$ for the rotation matrix $\rotmatrix(\angvec_i)$. 
The system's dynamics now includes the center-of-mass momentum $\momvec_i$ and angular momentum $\angmomvec_i$, such that the time evolution of each particle fulfills a Langevin equation ~\cite{martinetz2018gas},
\begin{equation}
\label{eq:langevin}
\begin{pmatrix}
\dot{\momvec}_i\\
\dot{\angmomvec}_i
\end{pmatrix} = - \dampingmatrix_i(\angvec_i) \begin{pmatrix}
\momvec_i\\
\angmomvec_i
\end{pmatrix} + \sqrt{2 \matrix{D}_i(\angvec_i)} \whitenoise_i(t) +  \begin{pmatrix}
\forcevec_i\\
\torque_i
\end{pmatrix}\,.
\end{equation}
 The first term in Eq.~(\ref{eq:langevin}) describes the friction of the surrounding environment, characterized by a damping $\dampingmatrix_i(\angvec_i)$, while the second term sets the diffusion of the system due to the thermal bath. 
 Together, both terms are responsible for the thermodynamical coupling to the environment. 
 The vector $\whitenoise_i$ describes an independent Gaussian white noise with variance 1 and the drag tensor $\dampingmatrix_i$ relates to the diffusion tensor through the fluctuation-dissipation relation, $\matrix{D}_i = \kB \tempenv \dampingmatrix_i \matrix{W}_i$ for $\matrix{W}_i = \begin{pmatrix}
\mass_i \matrix{1}& \matrix{0}\\
\matrix{0}& \matrix{\moi}_i\\
\end{pmatrix}\,.$
Finally, the last term in Eq.~(\ref{eq:langevin}) describes the force $\forcevec$ and the torque $\torque$ applied by the cooling and trapping fields.

\subsection{Torque-rotation correlation}
\label{app:torquerotation}
In Eq.~(\ref{eq:focus}), we expressed the correlation between the optical force applied by the cooling field and the motion of a spherical particle. 
Here, for non-spherical particles subject to an optical torque, we define the weighted torque-angular velocity correlation as,
\begin{equation}
\label{eq:focus2}
\corr_w^{\text{rot}} \coloneqq \frac{\sum_i \angmomvec_i \cdot (\matrix{\moi}_i^{-1} \torque_{\cool,i}) }{\sqrt{\sum_i \angmomvec_i \cdot (\matrix{\moi}_i^{-1} \angmomvec_i)}\sqrt{\sum_i \torque_{\cool,i} \cdot (\matrix{\moi}_i^{-1} \torque_{\cool,i})}}\,,
\end{equation}
which measures the co-linearity between the torque and the instantaneous rotation of the particle. 
By combining the rotational and translational contributions, we define the total weighted correlation as
\begin{equation}
\begin{split}
    \corr_w \coloneqq& \left(\sum_i \momvec_i^{\,2} \mass_i^{-1} + \sum_i \angmomvec_i \cdot (\matrix{\moi}_i^{-1} \angmomvec_i) \right)^{-1/2}\\
    &\left(\sum_i \forcevec_{\cool,i}^2 \mass_i^{-1} + \sum_i \torque_{\cool,i} \cdot (\matrix{\moi}_i^{-1} \torque_{\cool,i})\right)^{-1/2}\\
    &\left(\sum_i \momvec_i \cdot \forcevec_{\cool,i} \mass_i^{-1} + \sum_i \angmomvec_i \cdot (\matrix{\moi}_i^{-1} \torque_{\cool,i})\right)\,.
\end{split}
\end{equation}

\section{Simulations and data analysis}
\label{app:implementation}
The numerical code to simulate the particles' dynamics is written in Python with heavy usage of the library NumPy. 
All simulations involve a two-dimensional rectangular waveguide with open leads on both sides. 
The scalar Helmholtz equation defining the propagation of the light field, $[\Delta + \omega^2 \mu \epsilon(x,z)]\psi(x,z) = 0$, is simulated in 2D using the finite elements method (Library NGsolve: https://ngsolve.org/ \cite{schoberl1997netgen,schoberl2014c++}). 
The waveguide walls impose a Dirichlet boundary condition on the wave and perfectly matched layers absorb the outgoing waves in the asymptotic region.
We simulate the time evolution using the Euler method, except for the trapping field, where a higher-order Adams–Bashforth method is necessary to achieve energy conservation.
Collisions between particles among themselves and with the waveguide walls are rare, but incorporated via elastic collisions. 
The force on a rigid body with a uniform dielectric constant is calculated by numerical integration over the surface of the particle $\partial V_i$, 
    \begin{equation}
        \forcevec_i = \frac{1}{4} (\epsilon - \epsilon_0) \int_{\partial V_i} |\psi|^2 \d \n\,
    \end{equation}
Rotations and torques are computed in a similar fashion. 

Throughout the simulation, the optical forces are updated every half cooling timestep, $\coolstep/2$, while gravity and the Langevin forces get updated every $\coolstep/10$. 
The scattering matrix is measured at a rate given by the sampling time $\coolstep$ and the time derivative is approximated by taking the finite difference. 
The trapping field (incoherent to the cooling field) is introduced by performing a second FEM simulation at the corresponding wavelength.

In section \ref{sec:trapped}, the power spectral density, $|S_{zz}|$, was calculated using an implementation of Welch's method in the python library SciPy with a flat top window. 
The angular frequency $\omega_z$ of the trap was identified by the highest peak in the power spectral density. 
After cooling, the center-of-mass temperatures, $T_{\text{eff}}$, of the different trapped particles were estimated through a fitting of their power spectral densities with a Lorentzian model \cite{Li_2011} reading
\begin{equation}
    |S_{\text{zz}}(\omega)| = \frac{2 \kB T_{\text{eff}} \damping_{\text{tot}} }{\mass [(\omega^2-\omega_z^2)^2 + \omega^2 \damping_{\text{tot}}^2]},
\end{equation}
in which $\damping_{\text{tot}}$ stands for the total damping rate (i.e., sum of the environmental damping rate and the damping developed by the cooling forces). 

\section{Cooling efficiency}
\label{sec:app:efficiency}
\subsection{Derivation of the cooled energy}
The decrease in total energy reachable through our cooling approach can be estimated analytically. 
Here, the derivation will be performed for the case of trapped particles but a similar expression can be derived for freely-moving objects. 
For simplicity, we start out with a single translational degree of freedom, $z$, and approximate the trapping through a linear restoring force. 
The dynamics of each particle fulfills,
\begin{equation}
\ddot{z} + \gamma \dot{z} + \omega_0^2 z = \sqrt{2 k_{B} \tempenv \damping \mass^{-1}} \xi(t) + \mass^{-1} \force_{\cool}\,,
\label{app:eq:dynamics}
\end{equation}
in which $\omega_0$ stands for the resonance frequency of the trap, $\gamma$ for the damping rate and $\xi$ for a white noise (see Eq.~(\ref{eq:langevin_translation}) of the main text). 
Here, we neglect the influence of gravity, which plays a marginal role owing to the strength of the optical trap.

First, we outline the approach taken here. 
For systems having reached the stationary cooled energy, $\Ecool$, we assume that the system is ergodic and we make use of the fact that the average energy shift over all realizations is zero, namely $\timemean{\Delta E} \approx 0$.
In the following, we examine the different contributions to the changes in energy, before we ultimately derive an analytic formula for the cooled energy.

\subsubsection{Energy shift}
We focus here on strongly underdamped systems for which $\damping \ll \omega_0$. 
In order to solve the differential relation of Eq.~(\ref{app:eq:dynamics}), we make use of its Green's function,
\begin{equation}
\begin{split}
G(t) &\coloneqq e^{-\gamma t/2} \frac{\sin(t\sqrt{\omega_0^2-\gamma^2/4})}{\sqrt{\omega_0^2-\gamma^2/4}} \Theta(t) \\
&= t \Theta(t) [1 + \O((\omega_0 t +\gamma/\omega_0)^2)]\,
\end{split}
\end{equation}
where $\Theta(t)$ stands for a Heaviside function. 
This allows us to obtain the solution $z(t)$ as a convolution with the driving terms [right-hand side of Eq.\eqref{app:eq:dynamics}].

Different cooling forces are applied at each sampling time and $\coolstep$ is assumed smaller than the time scale of the particles' oscillations within the trap (i.e., $\coolstep \omega_0 \ll 1$). 
We start out by looking at the stochastic averaged gain over a single timestep. 
We identify a stochastic and a deterministic contribution ($\EX{\Delta E_{\tot}} = \Delta E_{\text{sto}} + \Delta E_{\text{det}}$), which, for a particle starting at position $z_0$ with momentum $\mom_0$, reads as follows 
\begin{equation}
\begin{split}
\Delta E_{\text{sto}} \coloneqq&  \frac{\EX{(\mom - \EX\mom)^2}}{2 \mass} +  \frac{\mass \omega_0^2 \EX{(z - \EX z)^2}}{2},\\
\Delta E_{\text{det}} \coloneqq&  \frac{(\EX \mom)^2 - (\mom_0)^2}{2 \mass} +  \frac{\mass \omega_0^2 [(\EX z)^2 - (z_0)^2]}{2}\,.
\end{split}
\label{app:eq:deltaE}
\end{equation}
In Eq.~(\ref{app:eq:deltaE}), $\EX{\cdot}$ marks the average over fluctuations imposed by the white noise $\xi$. 
For $z-\EX{z} = \sqrt{2 k_{B} \tempenv \damping \mass^{-1}} G * \xi$, where $G * \xi$ is the convolution of $G$ and $\xi$, the stochastic term in Eq.~(\ref{app:eq:deltaE}) reads,
\begin{equation}
\label{app:eq:Stoch}
\Delta E_{\text{sto}} = \kB \tempenv \gamma\coolstep \left[1 + \O((\damping/\omega_0 + \omega_0 \coolstep)^2) \right]\,.
\end{equation}
We will now average over the initial conditions $\mom_0,\pos_0$ (marked as $\initmean{\cdot}$). 
Due to the fact that the stochastic-energy changes do not depend on initial conditions, we get $\initmean{\Delta E_{\text{sto}}} = \Delta E_{\text{sto}}$. 
Going forward, we will make the assumptions that the initial position $\pos_0$ has mean $0$ and that it is uncorrelated with the force and momentum, namely $\initmean{\forcecool \pos_0}=\initmean{\mom_0 \pos_0}=0$.
Before looking at the deterministic term in Eq.~(\ref{app:eq:deltaE}), we note that $\EX{z} = z_0 \cos(\sqrt{\omega_0^2-\gamma^2/4}t) + \mom_0 \frac{\sin(\sqrt{\omega_0^2-\gamma^2/4}t)}{\mass \sqrt{\omega_0^2-\gamma^2/4}} + G * \mass^{-1} F_{\cool}$. 
The deterministic energy change averaged over all initial configurations reads,
\begin{equation}
		\label{app:eq:Det}
\begin{split}
 &\initmean{\Delta E_{\text{det}}} \approx \frac{2 \initmean{\mom_0 F_{\cool}}\coolstep + \initmean{F_{\cool}^2} \coolstep^2 }{2 \mass}\,,
\end{split}
\end{equation}
up to order,
\begin{equation}
\label{app:eq:higher_orders}
    \begin{split}
        &(\omega_0 \coolstep + \damping/\omega_0 )^2 M \times\\
        &\initmean{[M^{-1} F_\cool \coolstep  + M^{-1}\mom_0 + \pos_0 \omega_0]^2}.
    \end{split}
\end{equation}

\subsubsection{Cooled energy}
We will now look at the system in a stationary state and assume that it behaves ergodically. 
Thus, the stochastic mean can be replaced by a time average $\timemean{\cdot} = \mean{\cdot}[\xi,\mom_0,\pos_0]$. 
Here, $\timemean{\cdot}$ corresponds to the average of a physical quantity over the timesteps--i.e., for the momentum it is given by
\begin{equation}
    \timemean{\mom} \coloneqq \lim_{N\rightarrow\infty} \frac{1}{N+1} \sum_{n=0}^N \mom(t_n)\,
\end{equation}
for $t_n = t_0 + n \coolstep$ and $t_0$ corresponding to the arrival time in the cooled configuration.
At this stationary state, the energy does not change on average and we thus have
\begin{equation}
    0 = \timemean{\Delta E} = \mean{\Delta E_\tot}[\xi,\mom_0,\pos_0].
\end{equation}
Combining this relation with Eqs.~\eqref{app:eq:Stoch} and ~\eqref{app:eq:Det} and replacing the stochastic average with the time average, we get
\begin{equation}
\begin{split}
\label{eq:singlepart:zeroenergychange}
0 \approx& \frac{2 \timemean{\mom F_{\cool}}\coolstep + \timemean{F_{\cool}^2}\coolstep^2 }{2 \mass}\\
&+\kB \tempenv \damping \coolstep.
\end{split}
\end{equation}
We will next generalize this result to many particles with multiple degrees of freedom and uncoupled friction coefficients. 
We will denote by $\momvec_i$ the momentum vector of the $i$th particle. 
By assuming that the forces, momenta and positions are uncorrelated between the particles, Eq.~\eqref{eq:singlepart:zeroenergychange} generalizes to
\begin{equation}
\label{eq:delta_E_mean}
\begin{split}
	0 \approx & \sum_{i=1}^\Nscat \frac{1}{2 \mass_i}\left(2 \timemean{\momvec_{i}\cdot \forcevec_{\cool,i}} \coolstep + \timemean{\forcevec_{\cool,i}^2} \coolstep^2 \right) \\
&+ \kB \tempenv \sum_{i=1}^\Ndof \damping_i \coolstep\,.
\end{split}
\end{equation}
To simplify Eq.~(\ref{eq:delta_E_mean}), we introduce the time-averaged weighted inner product,
\begin{equation}
\inner[\vec{u}][\vec{v}][w] \coloneqq \timemean{\sum_{i=1}^\Nscat \frac{\vec{u}_i \vec{v}_i}{2\mass_i}}
\label{app:eq:innerprod}
\end{equation}
with a corresponding norm $\norm[\cdot][w]$.
Using Eq.~(\ref{app:eq:innerprod}), we can rewrite Eq.~(\ref{eq:delta_E_mean}) as,
\begin{equation}
\begin{split}
0 \approx& \,  2 \inner[\momvec][\forcevec_{\cool}][w]\coolstep  + \norm[\forcevec_{\cool}][w]^2 \coolstep^2 \\
&+ \kB \tempenv \Ndof \dofmean{\damping}\coolstep\,,
\label{app:eq:relation2}
\end{split}
\end{equation}
where $\norm[\forcevec_{\cool}][w]^2 = \inner[\forcevec_{\cool}][\forcevec_{\cool}][w]$ and $\dofmean{\cdot}$ is the average over the degrees of freedom.
By $E_{\text{cool}}^\text{\kin}$ we denote the kinetic energy of the system when the cooled steady state is reached.
Next, we note that in such a steady state, we have $\norm[\momvec_{\cool}][w]^2 = E_{\text{cool}}^\text{\kin}$ and we define the time-averaged weighted force-velocity correlation as,
\begin{equation}
\label{eq:meancorr}
\meancorr_w \coloneqq \frac{ \inner[\momvec][\forcevec_{\cool}][w] }{\norm[ \momvec ][w] \norm[ \forcevec_{\cool}][w]}\,.
\end{equation}
For identical spherical particles, $\meancorr_w$ [defined in Eq.~(\ref{eq:meancorr})] can be identified as the time-averaged force-velocity correlation,
\begin{equation}
\meancorr_w \coloneqq \frac{ \timemean{\sum_i \momvec_i \cdot \forcevec_{\cool,i} }}{\sqrt{\timemean{\sum_i \momvec_i^{\,2}}}\sqrt{\timemean{\sum_i \forcevec_{\cool,i}^2}}}\,.
\end{equation}

With this in mind, solving Eq.~\eqref{app:eq:relation2} leads to
\begin{equation} 
\label{eq:endenergy_app}
E_{\text{cool}}^\text{\kin} = \left(\frac{\norm[ \forcevec_{\cool}][w]^2 \coolstep + \Ndof \kB \tempenv \dofmean{\damping} }{ 2\norm[ \forcevec_{\cool}][w]\meancorr_w}\right)^{2}\,.
\end{equation}
For freely-moving particles (i.e., without optical trap), Eq.~\eqref{eq:endenergy_app} provides the total cooled energy. 

In the case of trapped particles, the potential energy contribution is given through the equipartition theorem and we get 
\begin{equation} 
\label{eq:totalendenergy_app}
\Ecool = 2 E_{\text{cool}}^{\kin}\,,
\end{equation}
Therefore, from Eqs.~(\ref{eq:endenergy_app}) and (\ref{eq:totalendenergy_app}) one can readily deduce Eq.~(\ref{eq:endenergy1}) of the main text.

\subsection{Analysis of the cooled energy}
\label{sec:app:endenergy}
The first term in the cooled energy given in Eq.~(\ref{eq:endenergy_app}) (referred to as the discretisation term) shows that at some point the momentum transfer provided by the cooling forces becomes too strong between steps.  The second term (referred to as the fluctuation term) corresponds to the 
inability of the algorithm to cool lower than the energy that gets added in each step due to white noise fluctuations. 

Next, we note that the cooled energy given in Eq.~(\ref{eq:endenergy_app}) is convex in $\norm[ \forcevec_{\cool}][w]$ and therefore a minimal cooled energy exists. 
The optimal cooled energy reads, 
\begin{equation}
\Ecool^{\text{opt}} = (\meancorr_w)^{-2} \Ndof \kB \tempenv \coolstep \dofmean{\damping}\,,
    \label{eq:endenergy_app_opt}
\end{equation}
and is obtained for an averaged weighted force,
\begin{equation}
    \norm[ \forcevec_{\cool}][w] = \sqrt{\frac{\kB \tempenv \Ndof \dofmean{\damping}}{\coolstep}}\,.
\end{equation}

\begin{figure}[t]
	\begin{tabular}{l{0.03\columnwidth} f{0.95\columnwidth}}
	(a)&%
	\hspace*{-10pt}
        \includegraphics[scale=1,valign=t]{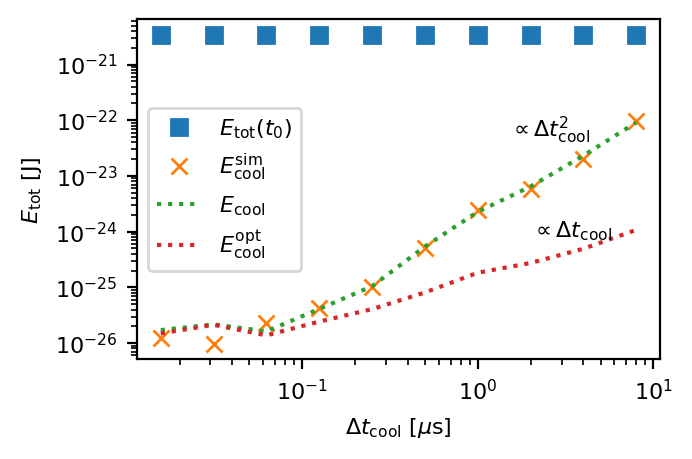}
		\tabularnewline
    (b)&%
	\hspace*{-10pt}
	\includegraphics[scale=1,valign=t]{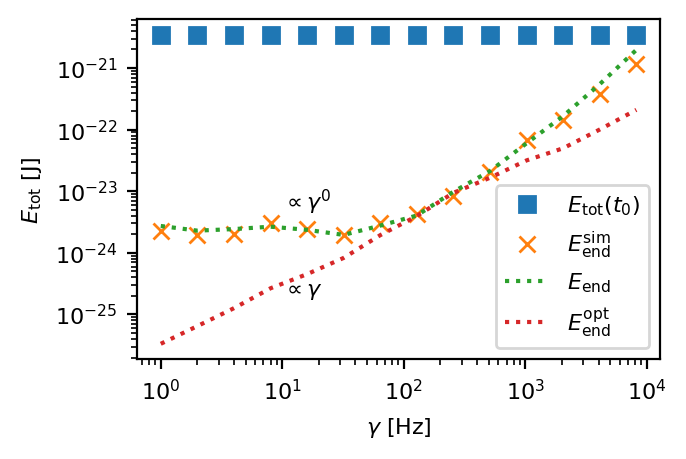}
	\end{tabular}
	\caption{(a), (b) Initial total energy (blue squares, $E_{\tot}(t_0)$) and end total energy (orange crosses, $E_\text{cool}^\text{sim}$), recorded when systematically cooling ($P= 2 $~$\mu$W) a random configuration of circular nano-beads  ($r=75 $~nm and $M=10$). 
	In (a), the sampling time, $\Delta t_{\cool}$, is varied while the damping rate is fixed to $\gamma = 6$~Hz. 
	In (b), the damping rate is varied while the sampling time is fixed to $\Delta t_{\cool} = 1$~$\mu$s. 
	The green dashed (red dashed) marks the prediction given by Eq.~(\ref{eq:endenergy_app}) (Eq.~(\ref{eq:endenergy_app_opt})), labelled as $\Ecool$ ($\Ecool^{\opt}$).}
	\label{fig:damping}
\end{figure}

Comparing Eqs.~(\ref{eq:endenergy_app}) and (\ref{eq:endenergy_app_opt}), we observe that, while $\Ecool$ varies quadratically with the sampling time for a strong enough power, $\Ecool^{\text{opt}}$ evolves only linearly with respect to $\Delta t_{\cool}$. 
This point is emphasized in Fig.~\ref{fig:damping}(a), in which cooling was performed for an initial random configuration of the nano-particles, with a fixed number of $\modes=10$ optical modes and while varying $\Delta t_{\cool}$. 
The blue squares (orange crosses) represent the initial (averaged end) energy at the start (end) of the process that we label $E_{\tot}(t_0)$ ($E_\text{cool}^\text{sim}$). 
The green and red dashed lines indicate the prediction given by Eq.~(\ref{eq:endenergy_app}) and (\ref{eq:endenergy_app_opt}) that are labelled as $\Ecool$ and $\Ecool^{\opt}$, respectively. Owing to $\Ecool$ scaling quadratically and $\Ecool^{\opt}$ linearly for high $\Delta t_{\cool}$, the cooled energy for a fixed power ($P= 2 $~$\mu$W) reaches an optimum for $\Delta t_{\cool} \approx 0.1$~$\mu$s. 
For sampling times below this value, the end energy settles at a constant value. 
Similarly, from Eqs.~(\ref{eq:endenergy_app}) and (\ref{eq:endenergy_app_opt}), we also observe different scalings with respect to the mean damping rate $\gamma$. 
Figure \ref{fig:damping}(b) reproduces an analysis similar to the one made in Fig.~\ref{fig:damping}(a) but in which $\gamma$ is varied (same color code in both figures). 
We observe that the cooled energy reaches the optimal cooled energy for a damping rate close to $\approx 100$~Hz and saturates for lower values due to the application of a force that is too strong.

\section{Parameter range}
\subsection{Power and damping}
\label{app:coolinglimits}
In this section, we derive the bounds given in Eq.~\eqref{eq:bounds}. 
As a measure for the usefulness of our cooling protocol, we introduce the requirement that the cooling performance is considered effective if our procedure achieves a minimum reduction of $50\%$ of the system's total energy, which reads
\begin{equation}
\Ecool < N_{\text{dof}} \kB \tempenv/4\,,
\label{eq:conditionapp}
\end{equation}
where $N_{\text{dof}} \kB \tempenv/2$ corresponds to the total thermal energy at equilibrium for a system with $\Ndof$ degrees of freedom.
Combining Eq.~(\ref{eq:conditionapp}) with Eq.~(\ref{eq:endenergy_app}) and under the assumption that $\meancorr_w^{-2} \coolstep \dofmean{\damping} \ll 1/2$, we get
\begin{equation}
	\frac{2 \meancorr_w^2}{ (\dofmean{\damping})^2 } > \frac{N_{\text{dof}} \kB T}{\norm[ F_{\cool}][w]^2} > \frac{\coolstep^2}{\meancorr_w^2}\,.
	\label{eq:Condition}
\end{equation}
In Eq.~\ref{eq:Condition}, $\dofmean{\damping}$ corresponds to the average damping rate over all degrees of freedom and serves to establish the operating boundaries displayed in Fig.~\ref{fig:numscat}(d) of the main text. 
The upper bound is reached if the power is too weak to counteract the gain of energy due to thermal coupling. 
On the other hand if the momentum transfer in each cooling step is too strong then the lower condition is violated.

\subsection{Sampling rate}
\label{app:lambda}

\begin{figure}[t!]
%\vspace*{10pt}
	\begin{tabular}{l{0.03\columnwidth} f{0.95\columnwidth}}
	(a)&%
	    \hspace*{-10pt}\includegraphics[scale=1,valign=t]{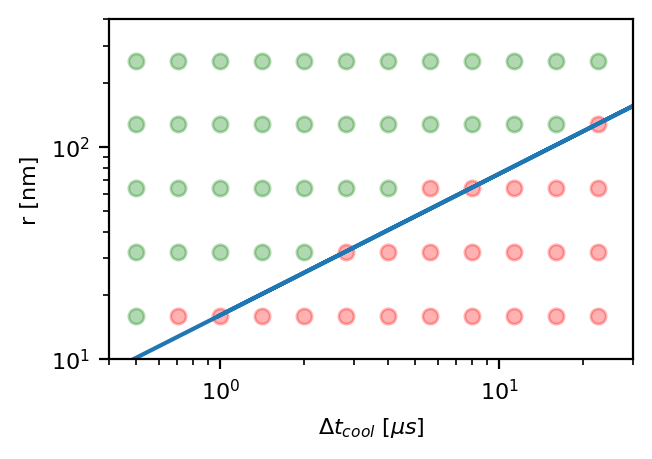}
		\tabularnewline
    (b)&%
	\hspace*{-10pt}\includegraphics[scale=1,valign=t]{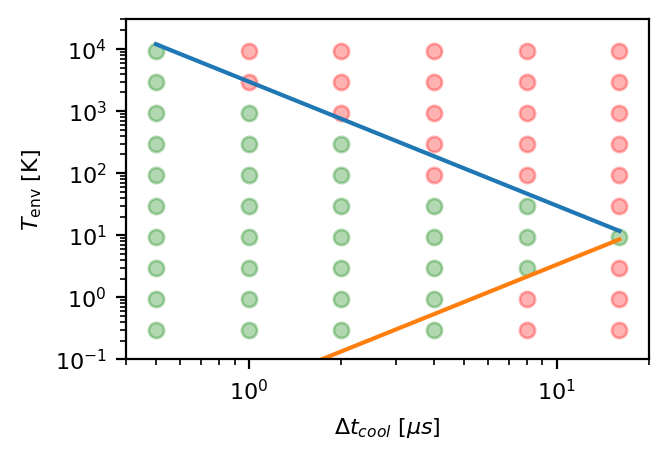}
	\end{tabular}
	\caption{(a) A system containing $\Nscat=10$ ciruclar nano-beads with various radii $r$ is cooled using different sampling times, $\Delta t_{\cool}$. 
	The green (red) dots indicate when the cooled energy is (is not) at most half the initial thermal equilibrium energy. 
	The blue line describes the boundary provided in Eq.~(\ref{eq:sampling}). 
	(b) Systems composed of  $\Nscat=10$ nano-beads are subject to different surrounding temperatures, $\tempenv$, and cooled using different sampling times, $\Delta t_{\cool}$. 
	The green (red) dots indicate when cooled energy is (is not) at most half the initial energy. 
	The blue line describes the boundary provided in Eq.~(\ref{eq:sampling}), while the orange line corresponds to the averaged lower boundary of Eq.~(\ref{eq:Condition}).
	In both cases good agreements between numerics and analytical estimates are found.}
	\label{fig:startenergyApp}
\end{figure}
In a simplified one particle model, it can be derived that the velocity magnitude $v$ needs to follow 
\begin{equation}
\label{eq:lambda_cond}
    v\Delta t_{\cool} < \lambda /4
\end{equation}
so that cooling can take place. 
Physically, Eq.~\ref{eq:lambda_cond} states that the average motion of the particles between two time measurements should remain small compared to $\lambda$ such that the corresponding modification of $\S$ can be tracked and counteracted in time.
With this in mind, we propose that cooling requires that the root mean square of the velocity over time follows the condition described in Eq.~\eqref{eq:lambda_cond}, namely $\sqrt{\timemean{\vec{v}_i^2}}\Delta t_{\cool} < \lambda /4$ for all particles $i$.

When the system resides in a steady state (and in absence of a potential) the total energy reads $\timemean{E_{\tot}}=\timemean{\sum_i \mass_i v_i^2/2}=\Ndof\kB \tempenv /2$, from which we can derive the relation,
\begin{equation}
\label{eq:sampling}
    \coolstep < \frac{1}{4}\sqrt{\frac{\sum_i \mass_i \lambda^2}{\Ndof\kB\tempenv}}\,,
\end{equation}
which establishes an operating criterion for the sampling time. 
For instance, Fig.~\ref{fig:startenergyApp}(a) displays the cooling effectiveness when our procedure is systematically applied on different initial configurations. 
Here, both the radius of the particles $r$ and the sampling time $\Delta t_{\cool}$ are varied. 
Green dots indicate when cooling was effective and the system energy is at least reduced by $50\%$. 
The red dots indicate when this condition is not fulfilled and we observe clearly the presence of a continuous range over which cooling operates. 
When the radius of the particles is changed, so is their mass ($m \propto r^3$). 
Thus, the condition of Eq.~(\ref{eq:sampling}) can be recast to relate $r$ and $\Delta t_{\cool}$, which is marked in Fig.~\ref{fig:startenergyApp}(a) by the blue line. 
In Fig.~\ref{fig:startenergyApp}(b) a similar analysis is performed for $\tempenv$ and $\Delta t_{\cool}$. 
The conditions of Eqs.~(\ref{eq:Condition}) and (\ref{eq:sampling}) are marked by the orange and blue lines, respectively, and again match nicely the simulations. 

\end{document}